\documentstyle[emulateapj]{article}
\begin{document}

\newcommand{\msun}{\mbox{${\cal M}_\odot$}}
\newcommand{\lsun}{\mbox{${\cal L}_\odot$}}
\newcommand{\kms}{\mbox{km s$^{-1}$}}
\newcommand{\HI}{\mbox{H\,{\sc i}}}
\newcommand{\mhi}{\mbox{${\cal M}_{HI}$}}
\def\hst{{\it HST}}
\def\etal{{\it et al.}}
\newcommand{\HII}{\mbox{H\,{\sc ii}}}
\newcommand{\am}[2]{$#1'\,\hspace{-1.7mm}.\hspace{.0mm}#2$}
\newcommand{\as}[2]{$#1''\,\hspace{-1.7mm}.\hspace{.0mm}#2$}
\def\lsim{~\rlap{$<$}{\lower 1.0ex\hbox{$\sim$}}}
\def\gsim{~\rlap{$>$}{\lower 1.0ex\hbox{$\sim$}}}
\newcommand{\iso}{sech$^{2}(z)$}

\title{Modelling the Interstellar Medium of Low Surface
Brightness Galaxies: Constraining Internal Extinction, Disk Color
Gradients, and Intrinsic Rotation Curve Shapes}

\vskip0.5cm
\author{L. D. Matthews\altaffilmark{1}}
\author{Kenneth Wood\altaffilmark{2}}

\altaffiltext{1}{National Radio Astronomy Observatory, 520 Edgemont Road,
Charlottesville, VA 22903 USA,
Electronic mail: lmatthew@nrao.edu}
\altaffiltext{2}{Smithsonian Astrophysical Observatory, 60 Garden Street,
Cambridge, MA 02138 USA, Electronic mail: kenny@claymore.harvard.edu}

\date
\singlespace
\tighten
\begin{abstract}
We use a combination  of 
three-dimensional Monte Carlo radiative transfer techniques
and  multi-wavelength ($BRHK$, H$\alpha$) imaging data to
investigate the nature of the interstellar medium (ISM) 
in the edge-on, low surface
brightness (LSB) galaxy UGC~7321. Using realistic models that incorporate
multiple scattering effects and clumping of the stars and the 
interstellar material,
we explore the distribution and  opacity 
of the interstellar material 
(gas+dust), and its effects on the observed stellar disk luminosity 
profiles, color gradients,
and rotation curve shape.  We find
that UGC~7321 contains a small but non-negligible dusty component
to its ISM, yielding a $B$-band optical depth $\bar\tau_{e,B}\sim$4.0
from disk edge to center. A significant fraction ($\sim50\pm10$\%)
of the interstellar material in the innermost regions of 
UGC~7321 appears to be contained in
a clumpy medium, indicating that LSB galaxies can support a modest,
multi-phase ISM structure.  In spite of the clear presence of dust,
we conclude that the large radial optical color gradients observed in
UGC~7321 and other similar LSB spiral galaxies 
cannot be accounted for by dust and must result primarily from significant
stellar population and/or metallicity gradients.
We show that realistic
optical depth effects will have little impact on the observed rotation curves
of edge-on disk galaxies and cannot explain the linear, slowly rising
rotation curves seen in some edge-on LSB spirals. Projection
effects create a far larger uncertainty in recovering the true underlying
rotation curve shape of galaxies viewed at inclinations $i\ga85^{\circ}$.

\end{abstract}


\section{Introduction}
\subsection{Background}
Low surface brightness (LSB) disk
galaxies can be broadly defined as rotationally-dominated 
galaxies with extrapolated face-on central surface
brightness at least $\sim$1 magnitude fainter than the canonical
Freeman (1970)
value of $\mu_{B,0}=$21.65~mag arcsec$^{-2}$. Although LSB galaxies
span the full range of Hubble types (e.g., Schombert \etal\ 1992), here we
use the term primarily in reference to the latest type spirals
(Scd-Sdm).  Numerous observational studies have
now shown that LSB galaxies are not ``abnormal'' objects, but rather are
a common product of disk galaxy formation and evolution. 
More extensive summaries of the properties of LSB
galaxies can be found in Bothun, Impey, \& McGaugh (1997) and Impey \&
Bothun (1997) and references therein. 

In spite of the vast body of 
multiwavelength observational
data now accumulated on LSB galaxies, several aspects of these systems  
still remain
enigmatic. One 
important example is the  detailed nature of their interstellar medium (ISM). 
Although the bulk of LSB galaxies are known to be rich in neutral
hydrogen gas (\HI), they are generally presumed to be 
dust and molecule-poor 
systems. Numerous observations directly and indirectly
support this picture, including: low metal
abundances (e.g., McGaugh 1994; R\"onnback \&
Bergvall 1995), low statistically 
inferred internal extinctions
(e.g., Tully \etal\ 1998; Matthews, van Driel, \& Gallagher 
1998), strong similarities between
optical and near-infrared (NIR) morphologies (Bergvall \etal\ 1999);
undetectable CO fluxes (Schombert
\etal\ 1990; Knezek 1993; de Blok \& van der Hulst 1998; but see Matthews \&
Gao 2000), 
and the high transparency of their stellar disks
(e.g., O'Neil \etal\ 1998; Matthews \etal\ 1999; Matthews, Gallagher,
\& van Driel 1999, hereafter MGvD99). Theoretical and numerical
models also suggest that the low surface densities and low
metallicities of LSB galaxies cannot support significant molecular gas
fractions (e.g., Mihos, Spaans, \& McGaugh 1999; 
Gerritsen \& de Blok 1999). Nonetheless, since most presently known
LSB galaxies are actively star-forming systems, having blue disk
colors and at least modest amounts of H$\alpha$ emission, it seems
likely that at
least some molecular material must be present in these systems.

Even if dust and molecular gas
contents in LSB galaxies are typically
small, extinction and reddening due to interstellar material may still have a
non-negligible impact on the observed properties of these
galaxies, and molecular gas could still be an important constituent of
the ISM (e.g., Spaans 1999).  For example, estimates of dust reddening, even
if small,
are needed to correctly interpret the stellar populations inferred from
broad-band colors and spectroscopic measurements (e.g., 
Bell \etal\ 2000). 
It has been found that many of the lowest-luminosity LSB
galaxies deviate from the standard optical
Tully-Fisher relation in the sense that they
rotate faster than predicted for their luminosities (e.g., Matthews,
van Driel, \& Gallagher 1998; Stil 1999; McGaugh \etal\ 2000), 
while more luminous LSB
galaxies seem to follow the same Tully-Fisher relation as higher surface
brightness galaxies (e.g., Zwaan \etal\ 1995; Sprayberry \etal\ 1995).
This appears to have important implications for galaxy formation models
(e.g., van den Bosch 2000) and for establishing the possible existence
of a ``baryonic'' Tully-Fisher relation (e.g., Matthews, van Driel, \&
Gallagher 1998; McGaugh \etal\ 2000); 
however, the internal extinction corrections appropriate
for the LSB galaxies used in these analyses remains a point of contention
(e.g., Han 1992; Rhee 1996; Matthews, van Driel, \& 
Gallagher 1998; Pierni 1999). Lastly, rotation curves 
of LSB galaxies offer hope to furthering our
insight into the nature of dark matter in galaxies (e.g., de Blok \&
McGaugh 1997; McGaugh \& de Blok 1998; Kravtsov \etal\ 1998; Swaters,
Madore, \& Trewhella 2000). 
Accurately interpreting high resolution optical rotation curve
data, which can play a crucial part in such analyses,  requires an
assessment of the role
of internal extinction in these systems (see Section~\ref{curves}).
In short, a more extensive and quantitative knowledge of the total amounts and 
distribution of dust and other molecular material
in LSB disks is overall vital to 
better understanding the global properties as well as the 
star-forming and evolutionary histories of these
systems.

To date, few attempts have been made to model the dust in
individual LSB galaxies in a sophisticated manner.
One difficulty is that 
these objects tend to be very
weak far-infrared and sub-millimeter sources (e.g., Hoeppe \etal\ 1994;
Pickering \& van der Hulst 1999;
Bell \etal\ 2000), hence direct
observations of re-emitted thermal radiation from their dust are
lacking.
Observing the dust in absorption at optical and near-infrared
wavelengths can also be problematic, since for  LSB
systems viewed at moderate or low inclinations, the signatures of dust can be
difficult to infer against  the patchy and diffuse
background light of the stellar disk. The data necessary to test dust
models are therefore sparse.
For these reasons, {\it 
edge-on} examples of
LSB galaxies are particularly valuable.

Because the
viewing geometries of edge-on LSB galaxies 
provide maximal path length through
the disks, such systems furnish us with the
unique opportunity to directly observe the vertical
and radial extent of the dust distribution, as well as any
associated dust-induced vertical and radial color gradients. 
With the aid of 
3D radiative transfer models, it becomes possible to
constrain the amount and distribution of dust and quantify its
effects on the observed
properties of the galaxy, including: total internal extinction, effects
on observed color  gradients, and effects on the disk rotation
curve. Properties of the disk viewed at other arbitrary inclinations
can also be predicted. Such techniques
have been used to explore the effects of dust in more luminous, high
surface brightness galaxies seen edge-on (e.g., Kylafis \& Bahcall
1987; Kuchinski \& Terndrup 1996; Kuchinski \etal\ 1998;
Trewhella, Madore, \& Kuchinski 1999; Xilouris \etal\
1997,1998,1999). Here we present one of the first applications
of full 3-D Monte Carlo radiative
transfer techniques to studying systems at the low optical depth regime of 
LSB galaxy disks.

\subsection{UGC~7321: A ``Superthin'' LSB Galaxy\protect\label{7321}}

UGC~7321 is a nearby example of an Sd spiral galaxy with an
extraordinarily thin, highly flattened and diffuse stellar disk with no
obvious bulge component (MGvD99; Matthews 2000).
MGvD99 showed that this ``superthin'' 
galaxy is actually an example of an LSB
spiral galaxy seen near edge-on ($i\approx88^{\circ}$). MGvD99
estimated that after correction for projection and internal
extinction,
UGC~7321 would have a $B$-band central disk surface brightness 
$\mu_{B,i}(0)\sim$23.6~mag arcsec$^{-2}$.

Although the {\it observed} surface brightness of UGC~7321 is enhanced
through projection,
several lines of evidence clearly indicate this is
an intrinsically LSB galaxy, independent of internal extinction
corrections. These include: the extreme transparency of
its disk  (e.g., several
background galaxies are clearly visible directly 
through it; MGvD99); the small stellar scale height and 
very low estimated stellar velocity dispersion (implying
a very low disk surface density; Matthews
2000); low emission line intensity ratios (Goad \& Roberts 1981);
weak far-infrared and radio continuum emission (see below);
and a rather high $M_{HI}/L_{B}$ ratio (1.1 in solar units). 

UGC~7321 is an ideal  LSB galaxy for detailed
modelling since it is nearby ($D\approx10^{+3}_{-3}$~Mpc; 
Gallagher \etal\ 2000)
and well-resolved.  Some key properties of UGC~7321
are summarized in Table~1. Although UGC~7321
is physically smaller, less massive, and less luminous 
than most other well-studied edge-on spirals,
UGC~7321 appears to be a prototypical example a relatively 
common class of late-type LSB
galaxies seen on edge (see also Goad \& Roberts 1981; Karachentsev,
Karachentseva, \& Parnovsky
1993; Bergvall \& R\"onnback 1995; Dalcanton \& Schectman 1996; 
Gerritsen \& de Blok 1999;
Matthews \& van Driel 2000). Possible face-on analogs of UGC~7321
might be such systems as NGC~4395 (see Sandage \& Bedke 1994) or
ESO~305-009 (see Matthews \& Gallagher 1997).

\subsubsection{The Data\protect\label{data}}
Much of the primary observational data we use for our present analysis were
described in MGvD99; these include near-infrared (NIR) $H$-band 
and optical ($B$- \& $R$-band) imaging data, and narrow-band H$\alpha$
imaging of UGC~7321. In addition, we supplement these
data with observations from several additional sources. {\it
Hubble Space Telescope} ({\it HST}) Wide Field
and Planetary Camera~2 $F702W$
and $F814W$ ($R$- and $I$-band equivalent) 
imaging observations of the inner disk
regions of UGC~7321 (Figure~\ref{fig:WFPC2}) are an important component
of our study and will be
discussed further in Gallagher \etal\ (2000).
We also used the optical longslit rotation curve data of 
Goad \& Roberts (1981), and NIR $H$ and $K_{s}$ images of UGC~7321 
obtained from the Two Micron All Sky Survey (2MASS) database
(http://sirtf.jpl.nasa.gov/2mass/). 

\subsubsection{Color Gradients in UGC~7321\protect\label{truegrad}}
One particularly intriguing finding of 
MGvD99 is that UGC~7321 exhibits
quite strong radial color gradients: 
$\Delta (B-R)\sim$1.0 magnitude (before correcting
for internal reddening), with the outer
disk regions being considerably bluer than the galaxy center
(Figure~\ref{fig:BRreal}). 
These authors
interpreted this as evidence of significant
stellar age gradients as a function of radius in the disk, and
suggested that this is a galaxy that is likely to have evolved relatively 
slowly, from the inside out. Large radial color gradients
[$\Delta (B-R)\ga 0.5$] were also
reported for a sample of 3 additional superthin galaxies by
Matthews, Gallagher, \& van Driel (2000). The
strength of the observed color gradients in UGC~7321 may also indicate 
that viscous evolution
has been minimal in this disk (cf. Firmani, Hern\'andez, \& Gallagher 
1996), leading
MGvD99 to postulate that galaxies like UGC~7321 may be some of the
most pristine star-forming disks in the nearby universe, and hence ideal
systems in which to explore disk evolutionary processes.

MGvD99 also reported significant {\it vertical} color gradients in the
UGC~7321 disk (as large as $\Delta (B-R)\sim$0.45, with the colors
reddening as a function of increasing $z$ height). Such gradients are
predicted to occur in galaxy disks as a result of dynamical heating
processes (e.g., Just, Fuchs, \& Wielen 
1996), but in practice they are difficult
to observe and interpret in most edge-on galaxies
due to the effects of seeing, contamination
from bulge light, and particularly due to contamination from
dust. Thus relatively few empirical constraints exist for present
models of vertical disk heating.

A critical assumption in interpreting the physical significance of the 
large color gradients observed 
in the UGC~7321 disk, as well as the significant gradients observed in
the disks of other less inclined
LSB spirals  (e.g., de Blok, van der Hulst, \& Bothun 1995; Bell \etal\ 2000)
is that reddening due to dust
is small. For example, in a sample of ``normal'' edge-on spirals, de
Grijs (1998) argued that to within observational errors, the 
observed color gradients and scale length
differences in different wavebands could be explained solely from
dust. 

Based on {\it HST}
imaging  observations, Matthews (1998) and
Gallagher \etal\ (2000) have shown that although UGC~7321, as well as
another nearby, edge-on LSB system UGC~711 lack the quintessential
dust lanes seen in brighter edge-on spirals (cf. Howk \& Savage 1999),
they are not dust-free.
Nonetheless, using a simple foreground screen model, MGvD99 suggested that
in the case of UGC~7321,
dust reddening could not explain more than $\sim$0.2 magnitudes of the
observed $B-R$ color gradient. Using both foreground screen and 
Triplex dust models (see Disney, Davies, \& Phillipps 1989), 
Bell \etal\ (2000) argued that the amount of dust reddening needed to
produce the observed color gradients in their sample of less inclined
LSB galaxies
was also too large to be consistent with observations.

To obtain a first estimate of the role of dust in UGC~7321, MGvD99
employed a simple foreground screen model. These 
authors sought not to derive a realistic dust model for UGC~7321, but
rather to test the possibility that a significant fraction of the observed
color gradient in UGC~7321 could be caused by dust. To put an upper limit
on the dust-induced color gradient, these authors  assumed
that the galaxy was effectively optically thin in $H$, and that the entire
observed $R-H$ radial
color gradient could be 
due to dust. 

A reanalysis of the $R-H$ color profile of UGC~7321 
(Figure~\ref{fig:RHreal}),
actually shows a significantly
larger color gradient ($\sim 0.9\pm0.1$ mag) than that
reported by MGvD99. This difference was traced to an error by MGvD99 in
extracting the $H$-band radial axis profile. However,
because MGvD99 were extremely conservative in their attribution of the
full $R-H$ color gradient to possible dust reddening, as we
further demonstrate here, the validity of
their final
conclusions remains unaffected. 

Although the arguments presented by MGvD99 and Bell \etal\ (2000)
regarding the dust reddening in their sample LSB galaxies
are expected to be for the most part robust, they
nonetheless relied on simplifying assumptions,
including the  use of 
smooth, idealized dust geometries and the neglect of scattering
effects. And because the color gradients observed in
UGC~7321 and some other LSB galaxies
are often found to be
 large compared with those typical in most normal surface brightness
spiral galaxies (cf. de Jong 1996), it is
important to establish  (1) whether galaxies like it are in fact a genre
of nearby disks with some of the largest intrinsic color gradients,
(2) whether these are pathological 
LSBs that contain more dust than average, and (3) whether our knowledge
of the dust in typical LSB galaxies is somehow incomplete. Detailed
studies of individual galaxies rather than statistical studies are
needed to address these questions.
If the large observed color gradients are
intrinsic (i.e., due to true stellar population and/or metallicity gradients)
rather than a result of dust reddening,
this is important not only for understanding  the evolution and stellar
contents of
these and other LSB 
disks, but also because such gradients may imply non-negligible
changes in the stellar mass-to-light ratio
($\Upsilon_{\star}$)  of disks as a function of radius. Such
trends could have an important consequence for mass-modelling of LSB
galaxies (i.e., for determining their dark matter halo parameters from
rotation curves; e.g., Kent 1986) 
and understanding the underlying physics behind the 
 Tully-Fisher relation 
(cf. Matthews, van Driel, \& Gallagher 1998; Stil 1999;
McGaugh \etal\ 2000). It is of interest therefore
to derive sophisticated, realistic dust models of UGC~7321 as a first 
test case.

We note that an 
advantage of studying color gradients in edge-on or near edge-on disks is
that due to the higher projected surface brightnesses of the outer
disk compared to a face-on system,
we can better observe the faintest (and perhaps least evolved) outer
disk regions. The trade-offs are
of course the uncertain 
effects of dust toward the central disk regions,
as well as the possible superposition of many different
stellar populations along a given line of sight. This makes the use of full
three-dimensional (3D) models critical for unravelling the effects of dust
on the observed colors and inferred stellar population
distributions. This is the approach we utilize in the present work.

\subsection{The Present Study}
The fundamental questions this paper seeks to address are:
\begin{itemize}
\item
What is the integrated optical depth of the dusty medium  in UGC~7321,
and what are its total mass and distribution?

\item What internal extinction corrections are appropriate for edge-on
LSB galaxies?

\item
How much of the observed color gradients in UGC~7321 and other
similar galaxies can be attributed to 
dust?

\item
What are the predicted dust-induced color gradients for analogous
LSB disks seen at alternate viewing angles?

\item
What is the impact of optical depth effects
on the shape of optically-derived rotation
curves of LSB galaxies observed near edge-on?

\end{itemize}

To answer these questions we use a suite of high resolution, 3D Monte Carlo 
radiation transfer codes that simulate scattered light images in
multiple wavebands and also can be used to model the 
effects of dust on optical rotation curves. As described below, 
our Monte Carlo radiation transfer code has recently been modified 
(see Wood \& Reynolds 1999 and Section~\ref{radtreat}) to enable the efficient 
simulation of 3D, optically thin systems, thus 
making it ideal for our present study of LSB galaxies. 
In doing this we also lay
the groundwork for future studies of the nature of dust in other LSB
galaxies. Much of the general framework required for realistically
modelling the effects of 
dust in galaxies has been discussed extensively by others 
(e.g., Kylafis \& Bahcall 1987; 
Bosma \etal\ 1992; Xilouris \etal\ 1997,1998,1999), including the
application of 
Monte Carlo methods to galactic environments (e.g., de Jong 1996;
Witt \& Gordon 1996,2000; Kuchinski \etal\ 1998; 
Bianchi, Ferrara, \& Giovanardi
1996; Bianchi \etal\ 2000). We do not attempt to repeat the 
detailed findings of these works here, but rather 
apply our codes to explore specific questions relating to LSB
galaxies. In particular, in this paper we focus on using our Monte
Carlo models to reproduce
the observed properties of the
edge-on LSB galaxy UGC~7321 and to predict the role dust may play on the
observed properties of similar LSB galaxies viewed over a range of
inclinations.  

\section{The Models}
The ingredients needed for our models of the dusty ISM of UGC~7321
are a 3D Monte Carlo 
radiation transfer code, measurements of 
dust and stellar distributions within the galaxy, 
and a characterization of 
the galactic velocity field.   The outputs 
of a Monte Carlo simulation are scattered light images
(Sect.~\ref{smooth} \& \ref{clumpscat})
and rotation curves 
that include the effects of internal
extinction (Sect.~\ref{curves}).
We now describe the model inputs.

\subsection{Radiation Transfer Treatment\protect\label{radtreat}}

In the present investigation, we 
perform radiation transfer calculations
with a 3D Monte Carlo scattering code that 
tracks photon energy packets as they are scattered and 
absorbed in a model galaxy.  The code is 
based on that described by Code \& Whitney (1995) and has been modified from 
the version
used for the smooth density and emissivity models of Wood (1997) and 
Wood \& Jones (1997) to enable us to study fully three dimensional systems 
(Wood \& Reynolds 1999).  
We construct the galactic density, emissivity, and velocity structure on 
a 3D linear Cartesian grid, and in the Monte Carlo radiation transfer, we now 
incorporate forced first scattering (Witt 1977) and a peeling off procedure 
(Yusef-Zadeh, Morris, \& White 1984).  This enables us to very efficiently 
generate images and rotation curves for the 3D, optically thin systems we are 
investigating (see also Wood \& Reynolds 1999). 
Unlike more simplistic dust treatments,
our code models both absorption and multiple scattering of photons;
both are necessary for accurately determining the total amount of
dust in a galaxy (e.g.,  Witt, Oliveri, \& Schild
1990; Block \etal\ 1994)  and for predicting its
observational manifestations.  

\subsection{Stellar Emissivity and Dust Distributions \protect\label{emiss}}

In Monte Carlo simulations of the transfer of starlight through galaxies,
the stellar sources are often represented by a smooth spatial distribution
(e.g., Wood \& Jones 1997; Ferrara \etal\ 1996, 1999; Bianchi, Ferrara, \&
Giovanardi 1996) rather than by individual point sources (but see recent
work by Cole, Wood, \& Nordsieck 1999; Wood \& Reynolds
1999). However, because
spiral disks seldom are observed to have completely smooth, uniform
disks at optical and NIR wavelengths, here we also consider the case
of starlight with an additional non-uniform component superposed on a
smooth background (see below).

In a purely smooth case, for a galaxy disk that is exponential in the radial
and vertical directions, the stellar emissivity can be expressed as
\begin{equation}
L(r,z)=L_{\lambda,0}{\rm exp}\left(-\frac{r}{h_{r}}-\frac{|z|}{h_{z}}\right)
\end{equation}

\noindent where $r$ and $z$ are the usual cylindrical 
coordinates, $L_{\lambda,0}$ is
the central disk surface density at a given waveband, and $h_{r}$ and
$h_{z}$ are the scale length and scale height, respectively, of the stars.
If a bulge is present, an additional term will be required in the emissivity
distribution.

A smooth density distribution of interstellar material (i.e.,
gas+dust), 
also assumed to be exponential in both
the radial and vertical directions, can be expressed as 
\begin{equation}
\rho(r,z)=\rho_0{\rm exp}\left(-\frac{r}{h_{r,d}}-\frac{|z|}{h_{z,d}}\right)
\end{equation}
where $\rho_0$ is the density (in g cm$^{-3}$) at the galactic center, and 
$h_{r,d}$ and $h_{z,d}$ are the scale length and scale height, respectively, 
of the dust. We then define an edge-on optical depth in the $B$-band,
$\tau_{e,B}$, measured from the disk edge to the disk center, as
\begin{equation}
\tau_{e,B}=\int\rho(r,0)\kappa_{B}{\rm d}r\approx \rho_{0}\kappa_{B}h_{r,d}
\end{equation}
\noindent where the absorption coefficient
$\kappa_{B}$ characterizes the opacity of the dust+gas mixture 
(see Sect.~\ref{grains}).

Because the ISM in real galaxies is not purely smooth,
several recent papers have examined the dust scattering of radiation in a
two-phase medium, with emphasis on the penetration and escape of
stellar radiation from clumpy environments (Boiss\'e 1990; Witt
\& Gordon 1996, 2000; Bianchi \etal\ 2000).  
In all of these studies, clumping
allowed photons to penetrate to greater depths than in a smooth medium, and
the escape of radiation becomes enhanced relative to the case where the same
gas and dust mass was distributed smoothly. The 
ability of Monte Carlo techniques to straightforwardly incorporate
these effects into models helps to make this technique particularly
powerful, hence we also explore the role of clumping in our
models in the present study.

For our study of the disk of UGC~7321 we
adopt the prescription for a 
two-phase interstellar medium from Witt \& Gordon (1996). Here the
medium consists of a smooth component 
and a clumpy medium. This
medium is characterized in terms of 
two parameters, namely the volume filling factor of dense
clumps, $ff$, and the density contrast between the clump and interclump
medium, $C$. $C$ in effect determines the total amount of mass
contained in the smooth versus the clumped ISM component.
We emphasize that in the work that follows,
such models generally contain {\it both} a
clumpy and a smoothly distributed ISM component; however, we hereafter
refer to these simply  as ``clumpy'' models.
 
We transform our smooth density distribution from Eq.~2 into a clumpy one
by looping through our 3D grid and applying the following
algorithm in each grid cell
\begin{equation}
n_{\rm clumpy}=\cases 
{n_{\rm smooth}/[ff+(1-ff)/C], & if $\xi < ff$;\cr
n_{\rm smooth}/[ff(C-1)+1],&otherwise.\cr}
\end{equation}
where $\xi$ is a uniform random deviate in the range (0,1).  This algorithm
assures that {\it on average} the total disk mass is the same for a
clumpy model as in a purely smooth model of a given optical depth, 
and that the ensemble average of the clumpy
distribution follows the same spatial profile as the smooth gas does.  In
this approach the smallest clump is a single cell in our density 
grid (Sect.~\ref{basicin}).

We have also applied the algorithm given by Eq.~4 to generate a
clumpy {\it emissivity} distribution, as could result, 
for example, from
clusters of young stars superposed on a smoother underlying disk of
older stars. 
We discuss our choices for $ff$ and for $C$ for both the stellar and
gas+dust distributions further below.

\subsection{Dust Opacity and Scattering Properties\protect\label{grains}}

We assume that the dust plus gas in our model galaxies is represented by a 
Kim, Martin, \& Hendry (1994) gas+dust mixture.
The Kim \etal\ models are an extension of the models of
Mathis, Rumpl, \& Nordsieck (1977) to include a larger distribution of grain
sizes, yielding a higher (and more realistic) estimate of the NIR
opacity. The wavelength 
dependent parameters defining this model are the total opacity,
$\kappa_{\lambda}$, 
scattering albedo, $a_{\lambda}$, 
and phase function asymmetry parameter, $g_{\lambda}$.  The 
scattering phase function is approximated by the Heyney-Greenstein phase 
function (Heyney \& Greenstein 1941).  In Table~2 we tabulate the adopted dust 
parameters for our simulations.

The total amount of dust in all of our models is characterized by the optical
depth parameter $\tau_{e,\lambda}$ (Eq.~3), 
which is the optical depth of the model
galaxy to a photon of wavelength $\lambda$  as it travels from the galaxy
center through the disk midplane to the observer.  The corresponding 
optical depth in the polar direction, through the galactic center, is 

$$\tau_{p,\lambda}=(h_{z,d}/h_{r,d})\tau_{e,\lambda}.$$

The Kim \etal\ (1994) gas+dust models that we employ  are
formulated to specifically represent {\it Galactic} grain properties,
and assumes a Galactic gas-to-dust ratio of 140. Because the gaseous
component of this ISM mixture contributes a negligibly
small fraction of
the total opacity, this assumption of a Galactic gas-to-dust ratio
will have no effect on  extinction or  reddening
effects derived from our models.   The actual
gas-to-dust ratio appropriate for UGC~7321 is discussed further
in Sect.~\ref{gtod}.

Evidence suggests that UGC~7321 may
be a rather metal-poor galaxy (Goad \& Roberts 1981), hence
it might be argued that  Galactic
grain and extinction properties
may not be applicable, and use of an alternate  extinction curve
[e.g., that from the Small Magellanic Cloud (SMC)] should be
considered.
However, we emphasize that
adoption of an SMC-type extinction curve (e.g., Bouchet \etal\
1985) 
would have no
significant effect on the results we present here. The Galactic and
SMC extinction curves are nearly identical in the optical, and begin
to deviate significantly only in the ultraviolet, i.e. at wavelengths
shortward of those considered in the present work (see, e.g.,  
Witt \& Gordon 2000). 

\subsection{The Model Galaxy: Basic Structural 
Parameters\protect\label{basicin}}
To constrain the total amount of dust opacity in a galaxy like UGC~7321,
we begin by attempting to construct models (composed of dust, gas, and
stars) that can reproduce the observed morphology of UGC~7321 in
various wavebands. Important constraints are also placed on our models by
the observed radial and vertical color gradients in UGC~7321, as
described in detail below.

We built our initial 
model galaxy based on the global disk parameters of UGC~7321
measured by MGvD99 and Matthews (2000): disk scale length 
$h_{r}$=2.1~kpc (as measured in $R$ band),  disk
scale height $h_{z}$=140~pc (as measured in $H$ band), and inclination
$i=88^{\circ}$.  For simplicity it 
was  assumed that these parameters have
no wavelength dependence. An axisymmetric,
exponential distribution was adopted along the radial and
vertical directions for both
the dust+gas and the stars, as in Eq.~1 \& 2. 
Using alternate analytic forms to describe the vertical light
distribution [e.g., a sech$(z)$ or sech$^{2}(z)$ function] would have no
appreciable effect on the conclusions that follow.

The scale
height  of the dust was assumed to be
one half that of the stars ($h_{z,d}=\frac{1}{2}h_{z}$; 
e.g., Evans 1994; Xilouris \etal\ 1997,1999), and the dust scale length was
taken to be $h_{r,d}$=1.5~kpc based on the observed dust clump
distribution in the WFPC2 imaging observations shown in 
Figure~\ref{fig:WFPC2}. Although a
number of workers have now argued for dust scale lengths equal to 
(e.g., Wainscoat, Freeman, \& Hyland 1989)  or
exceeding those of the stellar light  (e.g., 
Xilouris \etal\ 1999), such dust distributions are
inconsistent with the observed distribution of dust in
Figure~\ref{fig:WFPC2} (see also MGvD99).

The resolution of our models  is set by the size of our density grid
($200^3$ cells). 
Although the full measured radial extent of
UGC~7321 in the $R$-band is $\approx$8.1~kpc, in order
to obtain good spatial
resolution we truncated our model
calculations at $R_{max}=\pm 3h_{r}=$6.3~kpc.
This yielded a spatial resolution of $\sim$60~pc/pixel  in our finite 
200$^{3}$ cubic model grid and provides a good match to the projected pixel
size of our $H$-band images ($\sim$53~pc) and to the clump
sizes of $\sim 30-100$~pc 
inferred from the WFPC2 data in Figure~\ref{fig:WFPC2} (see also
Gallagher \etal\ 2000).

Because
UGC~7321 is essentially a pure disk galaxy (see MGvD99), no bulge
component was included. Also,
no spiral structure was added to our present models. Although
we cannot know the true face-on appearance of UGC~7321, several
arguments suggest that it is unlikely to exhibit well-defined spiral
structure (see MGvD99).
Moreover, to first order, the effects of spiral structure are  averaged
over in edge-on galaxies (see Xilouris \etal\ 1997), and
Misiriotis \etal\ (2000) have shown via detailed models
that derived dust and extinction parameters do not
change appreciably if spiral structure is included.

\section{Scattered Light Images and Color Gradients\protect\label{resultat}}
As described above, we tested models for UGC~7321 using  both
clumpy and smooth dust and stellar emissivity distributions
in order to assess the amount and distribution
of dust that can best reproduce observations of the galaxy. In this
section we examine the scattered light images, luminosity profiles, 
and color gradients predicted by various models, and compare them 
to the real observations.

\subsection{Smooth Models: Inputs and Results\protect\label{smooth}}
We began our modelling of UGC~7321 by generating a grid of models over
a wide range of optical depths, and having purely smooth stars and
dust. Because the smooth models are simpler (i.e., require fewer input
parameters) than the clumpy models, these allow us to zero in on a suitable
range of optical depths for modelling a diffuse, LSB 
galaxy like UGC~7321. Here we further examine four of these models, having
edge-on $B$-band optical depths of $\tau_{e,B}$=0.4, 2.0, 4.0 and 8.0
respectively. In Figure~\ref{fig:Rscat} (left), we show the 
$R$-band scattered light images corresponding to these models.

An examination of Figure~\ref{fig:Rscat} 
shows that we appear to have bracketed a suitable optical depth regime
for modelling UGC~7321, but that our smooth models
offer a relatively poor match to UGC~7321 morphologically. The
models overall lack the patchy light appearance of the real galaxy,
and in addition, the
$\tau_{e,B}\ge$4.0 models  exhibit clear dust lanes in the $R$-band
not seen in the real data. Even the $\tau_{e,B}$=2.0
model shows a brightness asymmetry about the midplane not seen in 
 $R$-band observations of UGC~7321. 
It is clear that in spite of the fact that an edge-on viewing angle
averages over many of the irregularities in the dust and stellar  
distributions, a purely
smooth model is insufficient to fully reproduce the observed properties of
UGC~7321.
Therefore in the next section consider models including a clumpy ISM
phase, as well as a clumped stellar component.

\subsection{Clumpy Models: Inputs\protect\label{clumpin}}
In this section we
investigate the effects of introducing a clumpy, 2-phase ISM as
well as a partially clumped stellar distribution on our Monte Carlo
models.

From our smooth models we have narrowed in on a range of optical
depths of interest for modelling 
UGC~7321 ($\tau_{e,B}=0.4-8.0$).  As a next step, we must determine
suitable values to use for the filling factor $ff$ and
density contrast parameter $C$ for both the stars and and the dust (see
Sect.~\ref{emiss}). 

We note  that for a clumpy medium,
the optical depth, $\tau$, 
is formally somewhat ill-defined (see Witt \& Gordon 2000). However,
because our clumpy models are formulated such that for a given value
of `$\tau$' they have
{\it on average} the same disk mass
as a smooth model with the same $\tau$ value (see Sect.~\ref{emiss}),
one can still characterize the medium in terms of what is effectively
a mean optical depth; we hereafter refer to this quantity as
$\bar\tau$. 

The clump filling factor for the ISM in real galaxies is very uncertain, and 
has been the focus of several other studies (e.g., 
Witt \& Gordon 1996; Bianchi \etal\ 2000). As examples,
typical values of around $ff=0.15$ were adopted by Kuchinski \etal\
(1998) and
Witt \& Gordon (2000), while values of 0.10-0.25 were chosen by Mihos, Spaans,
\& McGaugh (1999).  However, due to the expected greater sparsity of
molecular clouds in UGC~7321 compared with a higher surface brightness
or Milky Way-like
spiral, we feel the use of 
a somewhat lower filling factor of denser clumps is warranted. After trial and
error, for the present case
we find a value of
$ff_{gd}=0.06$ to be suitable.

The parameter $C$ for the ISM is also somewhat uncertain. 
%Witt \& Gordon (2000) used $C=$100 in their study of radiation
%transfer in a clumpy medium, while Mihos,
%Spaans, \& McGaugh (1999) suggested somewhat 
%lower values of $C$ (20 or 60) would be appropriate for the Galaxy. 
We therefore tested models
with $C$=100,40,20,9, \& 2 in order span the range of
values suggested by other workers for different environments, from
Milky Way-like galaxies (e.g., Witt \& Gordon 1996; Mihos, Spaans, \&
McGaugh 1999) to LSB galaxies (e.g., Mihos,
Spaans, \& McGaugh 1999).
Our choices of $C$ correspond to ratios of clumpy to smooth mass of
86:14, 72:28, 56:44, 36:64, \& 25:75, respectively.
In the present case, we found the best qualitative matches to the
data with  $C$=20. Models with less than $\sim$50\% of the opacity in
a clumped medium produced image morphologies too smooth to match the
data, while clumpier models tend to produce an overly mottled morphology.

Having established
 estimates of the ISM parameters, we first ran a grid of models for
various optical depths
containing a clumpy ISM but purely smooth starlight. Once again, the
resulting model images appeared too smooth to match the real
data. Therefore these models are not discussed further here.
Not surprisingly, we find that a partially clumped stellar distribution in
combination with a multi-phase ISM will be required to realistically
model UGC~7321.

To model a clumped emissivity distribution,  
we must now also choose appropriate values 
of $C_{*}$ and $ff_{*}$
for the stellar component of our models. 
In this case, we chose several values of $C_{*}$ in the
range representing rough approximations to the  total amount of
starlight from star-forming regions versus
smooth underlying disk regions that one might expect to see in  real LSB
galaxies.
We then ran grids of models with different stellar filling factors $ff_{*}$
until we found good matches to the observed image morphologies 
of UGC~7321 in the
$B$, $R$, and $H$ bands. We viewed each of our models from both an
$i=88^{\circ}$ and a
face-on orientation to check that our model looked qualitatively 
like a real late-type LSB galaxy disk from multiple viewing angles. 
Through this scheme we arrived
at preferred values of $ff_{*}$=0.60 and $C_{*}$=2.  This results in an
apportionment of $\sim$75\% of the light into higher density component and
$\sim$25\% into a smooth background component. However, we
emphasize that none of
our key conclusions regarding the total amount of dust in UGC~7321 or
its corresponding effects on the observed properties of the galaxy
are strongly sensitive to the choice of these values. 

Lastly, we note that
one might expect some small wavelength dependence on $C_{*}$ and
$ff_{*}$ to
account for the fact that generally
the starlight in galaxies becomes intrinsically smoother
as one moves from the optical to the NIR regime. 
However, as shown by Bergvall \etal\ (1999), LSB
galaxies tend to exhibit relatively minor morphological differences
between the optical and NIR compared with brighter spirals.  
Therefore we have settled
on one value of $ff_{*}$ and of $C_{*}$ that appear to give
satisfactory matches to the data over the range of wavelengths
considered here.

\subsubsection{Scattered Light Images from the Clumpy 
Models\protect\label{clumpscat}}
$R$-band scattered light images for our grid of clumpy models with
$\bar\tau_{e,B}$=0.4, 2.0, 4.0, and 8.0 are shown in
Figure~\ref{fig:Rscat} (right), alongside the smooth models from
Sect.~\ref{smooth} for comparison. It is immediately clear that
unlike the smooth models,
our clumpy models do not exhibit  dust lanes within this range of
optical depths.  Rather one sees the signatures of clumpy,
irregular absorption
over the central few kpc of the galaxy, consistent with the real
UGC~7321. All models also show a much less pronounced asymmetry about
the midplane compared with the smooth models of the corresponding
optical depth. In addition to the models shown in
Figure~\ref{fig:Rscat}, we also ran clumpy models with $\bar\tau_{e,B}$=12
and 24, but these models showed clear dust lanes and could be
immediately ruled out as having too much opacity.

Of the models shown in Figure~\ref{fig:Rscat}, 
the $\bar\tau_{e,B}$=4.0 case appears to
provide the best fit to 
multiwavelength observations of UGC~7321 on a purely morphological basis.
 In the $\bar\tau_{e,B}$=8.0
model, we see too few bright emitting regions more than one pixel
across, and somewhat too much central opacity. In contrast, the
$\bar\tau_{e,B}=0.2$ and 2.0 models both show stronger central
concentrations of starlight at $R$-band 
than in the real galaxy, and less patchiness
within the inner few kpc. The 
$\bar\tau_{e,B}$=4.0 has a 
discernible but weak central concentration of starlight, and little
evidence for absorption outside roughly $\pm$3~kpc, both in excellent
agreement with UGC~7321. The match of the
$\bar\tau_{e,B}$=4.0 model is also quite good at $H$-band, where it nicely
reproduces the more well-defined brightness center of the galaxy and
overall smoother appearance of the central regions of the galaxy as
seen at these wavelengths. In Figure~\ref{fig:zoom} we show more
detailed  close-ups of the
$\bar\tau_{e,B}$=4.0 $R$ and $H$ 
scattered light images compared with view of the real galaxy at the
corresponding wavelengths.

\subsection{Color Gradients Predicted from Clumpy 
Models\protect\label{clumpgrad}}
We now explore the color gradients predicted by our
models and 
compare them to those seen in  the real data.
In Figure~\ref{fig:smoothcolor} 
we plot the $B-R$ and $R-H$ radial color gradients along
the major axis extracted from both our smooth models (from
Sect.~\ref{smooth}) and our
clumpy models, for the optical depths
$\tau_{e,B}$=0.4, 2.0, 4.0, and 8.0. As expected, the color profiles
extracted from the clumpy models appear somewhat
``noisier'' than those from the smooth models. 
In fact, the intrinsic rms noise inherent in our models in 
only $\sim$0.005 magnitudes, hence the 
visible fluctuations 
in the radial color profiles in the lower portion of 
Figure~\ref{fig:smoothcolor} are real. 

From Figure~\ref{fig:smoothcolor} we see that at a given optical depth,
the total dust-induced color gradient (from disk edge to center)
is  virtually identical to that in
the smooth models. Likewise, the {\it mean} amount of reddening at a
given galactocentric radius is nearly identical in the two sets of
models; along a given sight line, differences appear to be $\le$0.015
magnitude. In general, the reddening effect of
a clumpy medium is expected to be smaller than that produced by a diffuse
model with the same amount of dust, particularly if the filling factor
of the clouds is small (e.g., de Jong 1996; Bianchi \etal\ 2000). 
However, our new models show that in the
low optical depth regime, even for the low clump filling factors used
here, this effect is nearly imperceptibly small, and that both the clumpy
and the smooth models produce roughly the same
amount of {\it mean} reddening, albeit with much larger statistical
variations between different lines of sight. We find this holds even
for a pure clumped medium with  no smooth component.

A further examination of
Figure~\ref{fig:smoothcolor} reveals that the observed
$\Delta(R-H)\sim$0.9 of Figure~\ref{fig:RHreal} cannot be reproduced
solely from dust effects 
for models with $\tau_{e,B}\le8.0$. Meanwhile, 
the $B-R$ color profiles for both the smooth and the clumpy models 
over this optical depth range show an additional interesting
behavior:  for $\tau_{e,B}\ge$4.0 the dust-induced $B-R$ color
actually {\it saturates} at $\Delta(B-R)\sim$0.31. In other
words, adding additional dust cannot further increase the total
dust-induced $B-R$ color
gradient to $>$0.31 magnitudes from the disk edge to the disk center.
As seen from 
Figure~\ref{fig:BRreal}, $\Delta(B-R)\sim$0.3 is significantly
less than the actual observed $B-R$ color gradient in UGC~7321 over
this radial interval, hence our 
models indicate that no amount
of dust can fully account for the observed $B-R$ color gradient in UGC~7321.

This saturation effect we see in the dust-induced
$B-R$ color gradients arises due to the wavelength-dependent
absorption in realistic systems where emission and 
absorption are mixed together (in contrast, for example, to
simple foreground screen models).  A simple illustrative example is a constant 
density absorbing medium (no scattering) with uniform emissivity,
i.e., a slab model where the stars and dust are spatially coincident.  The
emergent 
intensity is $I_\lambda\propto [1-e^{-\tau_\lambda}]/\tau_\lambda$.  From 
Table~2, we see that (approximately) $\tau_B:\tau_R:\tau_H = 8:4:1$, giving
\begin{equation}
{{I_B}\over{I_R}} \propto {{[1-e^{-\tau_B}]}\over {2[1-e^{-\tau_B/2}]}}\; ,\;\;
{{I_R}\over{I_H}} \propto {{[1-e^{-\tau_B/2}]}\over {4[1-e^{-\tau_B/8}]}}\; .
\end{equation}
Plotting $(B-R)$ and $(R-H)$ against $\tau_{B}$
for this simple 
example (Figure~\ref{fig:sat}), we see that 
$(B-R)$ saturates before $(R-H)$ due to the smaller opacity difference in 
going from $B$ to $R$, than in going from $R$ to $H$. The exact
optical depth at which the saturation occurs is slightly different for
this simple model than the more realistic models we have
computed. However it still demonstrates that  
even if we have significantly underestimated the amount of opacity in
UGC~7321's disk, our conclusion remains unchanged that
dust alone still cannot account for the large
color gradients
observed in UGC~7321.

Above we have argued that models for UGC~7321 with
$\bar\tau_{e,B}\ge$8.0 can be ruled out on a purely
morphological basis. To further demonstrate that we have not grossly
underestimated the amount of internal extinction in UGC~7321, 
we have also examined the $H-K$ color gradients. In Figure~\ref{fig:HKmodcolor}
we show the $H-K$ color profile along the major axis for our family of
clumpy models, and in   Figure~\ref{fig:HKreal} we
show a radial $H-K_{s}$
color plot for UGC~7321
derived from 2MASS data. The field-of-view of the available 2MASS
data is limited to $\sim200''$, and due to the large angular size of
UGC~7321 compared with the scan size of the survey, the data have very poor
signal-to-noise outside the central arcminute or so of the galaxy (see
Jarrett \etal\ 2000 for an description of 2MASS extended source data).
Nonetheless, the 2MASS
data are consistent with a $H-K$ radial color gradient of no more than
$\sim$0.1 magnitude in UGC~7321. 
Comparing this to our models, 
we see that $\bar\tau_{e,B}$=8.0
models produce a $\Delta(H-K)$ in excess of that expected on the basis
of the  2MASS results.

As a final test on our estimates of $\tau$, we lastly examine 
the {\it vertical}
$B-R$ color gradients (i.e., those parallel to the minor axis) 
for the same four optical depths as above (Figure~\ref{fig:fakevert}) 
and compare these with the
real data (Figure~\ref{fig:realvert}). 

In the real UGC~7321 data, near the minor axis,
the vertical $B-R$ color profile is seen to be redder by $\sim$0.31
magnitudes at $z$=0.15~kpc compared to  the highest observable
$z$-heights (Figure~\ref{fig:realvert}).  
However, we see that the
$\bar\tau_{e,B}$=8.0 model predicts a reddening of several tenths of a
magnitude higher than this
along the minor axis. Meanwhile, since reddening
toward the midplane along the $z$ direction is in general due to dust
rather than stellar population effects, 
it appears $\bar\tau_{e,B}>$2.0 is required in order to explain the
amount of reddening that is observed. Hence we again conclude that the
$\bar\tau_{e,B}$=4.0 model produces the best overall match to the data.

Near a
galactocentric radius of \am{1}{0} ($\sim$3~kpc), the $\bar\tau_{e,B}=4.0$
model predicts a dust-induced vertical $B-R$ color gradient  
of $\sim$0.15 magnitudes. In the real UGC~7321 data,
MGvD99 actually observed a {\it bluing} of $\Delta(B-R)\sim$0.3  toward
the galaxy midplane at this radius. Such a bluing toward low $z$-heights
is predicted to occur as a consequence of new young stars 
continually being born
near the galaxy midplane, while older stars are scattered to larger
$z$ heights via dynamical heating processes (e.g., Just, Fuchs, \&
Wielen 1996). 
The small amount of dust reddening helps to explain why UGC~7321
is one of the few galaxies studied so far 
in which we can  readily observe the predicted
stellar population changes with $z$-height. This demonstrates that edge-on LSB
galaxies like UGC~7321 can provide excellent laboratories for further
constraining dynamical heating processes.

In summary, we conclude that {\it a clumpy ISM model
with $\bar\tau_{e,B}\sim$4.0 provides the best characterization of the observed
properties of UGC~7321, and that dust alone cannot account for the large 
radial color gradients observed in this galaxy}. 
This suggests that 
the existence of significant stellar population gradients (and
possibly metallicity gradients) in the disk of UGC~7321, and
the likelihood that galaxies like UGC~7321
maintain a well-preserved record of how galaxy
disks have been built up  over time.

\subsection{Radial Intensity Profiles\protect\label{radplot}}
In Figure~\ref{fig:radprofs} we show $B$, $R$, and $H$
radial brightness profiles extracted along the major
axes for our grid of clumpy models. These profiles permit estimates of
the expected extinction along different lines of sight and in
different wavebands for a galaxy like UGC~7321. These profiles confirm
that for a $\bar\tau_{e,B}$=4.0 model, extinction effects are nearly
negligible in $H$-band, 
even at small galactocentric radii (see also MGvD99; Matthews 2000). 
They also show that in
$B$-band, extinction toward the the center of UGC~7321 should be
roughly 0.65$\pm$0.2 magnitudes in $B$-band and 0.43$\pm$0.2 
magnitudes in $R$, in
good agreement with the estimates made by MGvD99.

\subsection{UGC~7321 Analogs at Other Inclinations\protect\label{analogs}}
We have seen from our models in the previous sections that 
UGC~7321 is dust poor, but not dust free, and  that
small but non-negligible corrections must be made for the total internal
extinction and reddening caused by dust when interpreting the colors,
total magnitudes, and intrinsic surface brightnesses of LSB galaxies
viewed at high inclination. But what about at lower inclinations?

An advantage of our 3D Monte Carlo simulations is the ability to
easily examine a given model at any arbitrary viewing angle. We can
therefore  use  the models described above to predict how
dust will affect the observed properties of LSB disk
galaxies seen at inclinations higher or lower than 88$^{\circ}$. 
Other authors
have previously used Monte Carlo simulations to explore the effects of dust on
the colors and global extinctions of less inclined galaxies (e.g., de
Jong 1996), but our high
resolution, high signal-to-noise simulations allow us
to probe the detailed effects of dust on the observed light and color
distributions in the very low optical depth regime
appropriate for LSB galaxies and also to investigate the effects of a
clumpy ISM. All of the
models we discuss in this section have stellar and dust scale parameters
identical to those used in the UGC~7321 models of the previous
sections (see also Table~2).

In Figure~\ref{fig:lowincsm} \& \ref{fig:lowinccl} we show the predicted
$R-H$ and $B-R$ major 
axis color profiles for our $\tau_{e,B}$=0.4-8.0 
smooth and clumpy models, respectively, at
inclinations $i=0^{\circ}$, 45$^{\circ}$, 
and 90$^{\circ}$. For comparison, we also
replot our $i=88^{\circ}$ models from Figure~\ref{fig:smoothcolor}.

At $i=90^{\circ}$, the observed trends are much as expected. For the
various optical depths, we see
small increases of order 0.1 magnitude in the amplitude of the
$R-H$ color gradients compared with the $i=88^{\circ}$ models,
and difference of $\sim0.05$ magnitude in $\Delta(B-R)$ compared with the
$i=88^{\circ}$ case. 
The effects at $i=0$ and $i=45^{\circ}$ are somewhat more
interesting, although primarily from a theoretical perspective. 
In cases of galaxies viewed at high inclinations,
light scattered by dust has only a small
effect on the emergent  light profiles, particularly in a clumpy
medium. Instead the observed color changes are
dominated by dust absorption effects. 
In contrast, for $i\le45^{\circ}$ and the 
small optical depths we consider here, we
can see directly from our models how the dominance of scattering versus
absorption changes across the face of the galaxy. In these cases,
dust scattering actually causes a
{\it bluing} of the light towards the 
galaxy center. For the  higher optical depth ($\tau_{e,B}\ge$2.0)
models, this trend is interrupted
by  a rather abrupt reddening at the smallest galactocentric radii.
These trends are easier to see in the smooth models 
(Figure~\ref{fig:lowincsm}).
However, note that
at both $i=0$ and $i=45^{\circ}$, the maximum amplitude
of these effects are only a few hundredths of a magnitude and 
therefore are predicted to essentially unobservable
in real LSB galaxies.  We
conclude that  at low and moderate
optical depths, dust will have no appreciable effect on the observed
radial color gradients of moderately inclined  galaxy disks,
consistent with the findings of de Jong (1996).  Likewise, internal
extinction corrections at optical and NIR wavelengths should be almost
negligible for the bulk of LSB galaxies viewed at low or moderate inclinations.

\subsection{Discussion}
\subsubsection{Gas-to-Dust Ratios and Total Dust Content\protect\label{gtod}}
In Sections~\ref{clumpscat} \& \ref{clumpgrad}, 
we concluded that among the models we have considered,
a clumpy, multi-phase ISM model with 
$\bar\tau_{e,B}\approx 4.0$
provides the best overall match to the observed properties of the
edge-on LSB galaxy UGC~7321. We now consider some additional
implications of this model.

As described in Sect.~\ref{emiss}, 
we have parameterized our models in terms of
the opacity of a gas+dust mixture. For a given optical depth, 
the total mass of material
in our model ISM can then be computed as
\begin{equation}
M_{ISM}=\int\rho{\rm
dV}= 
\end{equation}
$$2\pi\rho_{0}\int_{-\infty}^{\infty}
\int_{0}^{R_{max}}\exp\left(\frac{-r}{h_{r,d}}-\frac{|z|}{h_{z,d}}\right)
{\rm d}z r{\rm d}r \approx 4\pi\rho_{0}h^{2}_{r,d}h_{z,d}$$
\noindent  From this, one can obtain the mass in the form of dust by
dividing Eq.~6 by the assumed gas-to-dust ratio of the absorbing
medium (in our case, $\rho_{g}/\rho_{d}$=140; see
Sect.~\ref{grains}).  
In terms of the edge-on optical depth $\tau_{e,B}$, the dust mass $M_{d}$
can thus be expressed in solar units  as 
\begin{equation}
M_{d}=2.2\times 10^{7} \cdot\left(\rho_{g}/\rho_{d}\right)^{-1}\tau_{e,B}~~M_{\odot}.
\end{equation}
\noindent For our $\bar\tau_{e,B}$=4.0 clumpy
models, this yields $M_{d}\sim 8.8\times10^{5}$~\msun, 
with 56\% of this in a clumpy
component, and the remainder in a smoothly distributed medium. 

{\it IRAS} far-infrared flux measurements are also available for
UGC~7321 (e.g., Sage 1993), allowing an independent estimate of the
dust content of UGC~7321. From the {\it IRAS} 100$\mu$m flux, one
can make an estimate of the mass of the 
far-infrared-emitting dust in a galaxy as:
\begin{equation}
M_{d,100\mu {\rm m}}=c_{g}S_{100}D^{2}(e^{144/T_{d}}-1) ~~M_{\odot}
\end{equation}

\noindent where $S_{100}$ is the {\it IRAS} 100$\mu$m flux density in Jy
(1.15~Jy for UGC~7321; Sage 1993), D is the distance in Mpc, $T_{d}$ 
is the dust
temperature, and $c_{g}$ is a
constant that depends on the grain opacity, and can vary by 
roughly a factor of two, depending on the composition of the grains
(e.g., Devereux \& Young 1990). Sage (1993) estimated
$T_{d}$=33.5~K for UGC~7321, and we adopt the value of $c_{g}$=5 suggested
by Thronson \& Telesco (1986). This yields
$M_{d,100\mu{\rm m}}\sim4.2\times10^{4}$~\msun\ for UGC~7321, or only $\sim$5\%
of the dust mass estimated from our models. However, Devereux \&
Young (1990) found from an examination of a large sample of {\it
IRAS}-detected nearby galaxies, that typically only $\sim$10-20\% of
the total dust mass in normal disk galaxies is warm enough to radiate in the
{\it IRAS} bands (see also Greenberg \& Li 1991; Whittet 1992).  Lisenfeld \&
Ferrara (1998) also found for a sample of dwarf galaxies
that the dust mass derived from
extinction estimates is on average $\sim29\times$ lower than that
derived from FIR data.
If this also holds for UGC~7321, then our estimated
dust mass from our models appears to correlate roughly as expected
with the estimate   from the
{\it IRAS} data, especially given the uncertainties in the {\it IRAS}
estimate (see, e.g., Lisenfeld \& Ferrara 1998).

Let us now return to the issue of the total interstellar 
gas mass implied by our
models. As described in Sect.~\ref{emiss}, our models are
parameterized in terms of a total opacity and assume
gas-to-dust ratio of 140 (Kim, Martin, \& Hendry 1994). As emphasized earlier,
although the exact  choice of the gas-to-dust ratio 
has negligible effect on the
extinction and reddening parameters derived from our models (since the
opacity of the gas is negligible), it does
however impact the total gas+dust  mass we infer: for 
$\rho_{g}/\rho_{d}$=140, we derive from Eq.~6 $M_{ISM}\approx 1.2\times 10^{8}$
for our $\bar\tau_{e,B}$=4.0 model.  

\HI\ spectral line measurements by MGvD99 yield a total integrated
\HI\ mass for UGC~7321 of \mhi=$1.1\times10^{9}$~\msun. Devereux \&
Young (1990) found that typically only the \HI\ gas
contained within roughly half of the optical radius of spiral galaxies
has appreciable dust
associated with it. From the pencil beam map of 
MGvD99, we estimate this corresponds to $\sim$50\% of
the total \HI\ content of UGC~7321, or $\sim$5.5$\times 10^{8}$~\msun. In
addition, Matthews \& Gao (2000) recently detected 
$^{12}$CO 1$\rightarrow$0 emission
from the central regions of UGC~7321, and estimated a molecular
hydrogen content within the central 2.5~kpc of the galaxy
of $\sim2\times10^{7}$~\msun\ using a standard
Galactic CO-to-H$_{2}$ conversion factor (but see below). Thus after
correction for helium, the total gas
content in the central regions of UGC~7321 is $\sim$6.2 times {\it
more} than the total ISM mass inferred from our models. This discrepency is
significant even after accounting for the range of accepted values 
for the Galactic 
gas-to-dust ratio [where $\rho_{gas}/\rho_{dust}\sim$120-140 based on 
depletion and extinction measurements (Whittet 1992)]. Not surprisingly,
{\it the assumption of a Galactic gas-to-dust ratio appears to be invalid
for UGC~7321.}

There is now considerable evidence that the gas-to-dust
ratio 
varies from galaxy to galaxy (e.g., Bouchet \etal\ 1985; Hunter \etal\
1989; Dwek 1998; Lisenfeld \& Ferrara 1998), with position within galaxies
(e.g., Stanimirovic \etal\ 2000),
and as a function of metallicity (e.g., Bouchet \etal\
1985; Lisenfel \& Ferrara 1998). For example,
Koornneef
(1982) found the gas-to-dust ratio for the Large Magellanic Cloud
(LMC) to be $\sim$4 times the Galactic value, while Bouchet \etal\
(1985) and Stanimirovic \etal\ (2000)
found a value for the SMC of $\sim$8-11 times Galactic (assuming a
Galactic gas-to-dust ratio of $\sim$140).

Although no oxygen abundance measurements are available for
UGC~7321, typical metallicity 
values for  LSB galaxies tend to be $\la 1/3$ solar
(e.g., McGaugh 1994; R\"onnback \& Bergvall 1995).  
That we require a gas-to-dust ratio of $\sim$6.2 times the Galactic value
to reconcile our model gas masses with observed values of
UGC~7321 thus seems appropriate
for a galaxy whose metallicity is likely to lie
somewhere between SMC and LMC values. The {\it global} gas-to-dust
ratio inferred from our extinction estimates 
and the total observed \HI\ content
of UGC~7321 (after correction for He) would then be $\sim$1600.

\subsubsection{Implications of a Clumpy, Multi-Phase ISM in 
UGC~7321\protect\label{clumpimp}}
We have shown that a dusty ISM with a significant fraction of
its total mass ($\sim$56\%)
in a the form clumpy component is needed to explain the observed
properties of UGC~7321. We now consider the implications of such a
finding.

For the type of simple, two-phase ISM model that we have used in the
present work, it is often assumed that the clumped component of the
medium should represent a predominantly molecular gas medium, while the diffuse
component represents diffuse atomic gas (e.g., Bianchi \etal\ 2000).
In the case of UGC~7321, if we make such assumptions, then we would predict a
molecular gas to atomic gas ratio somewhat
higher than the typical upper limits of $\la$40\% found for
the central regions of 
Sd-Sdm galaxies by Devereux \& Young (1990). Moreover, this is
considerably higher than empirical
measures for UGC~7321 by Matthews \& Gao (2000), which suggest that only
$\sim$4\% of the total gas content in the central regions of UGC~7321
is molecular (assuming a Galactic CO-to-H$_{2}$ conversion factor).

There are several possible explanations for this
trend. Firstly,  the association of the atomic gas solely to a diffuse
component, and the molecular gas solely to the clumped component in a
gross oversimplification of the ISM of any galaxy (e.g., Turner
1988; Elmegreen
1993). This may be particularly true in the interstellar environments
of LSB galaxies. The optical depths we infer for the ``dark clouds''
in UGC~7321 ($A_{V}\la$1.5; see also Gallagher \etal\ 2000) 
are consistent with the diffuse
molecular clouds, and the 
lowest density Giant Molecular Clouds in the Milky Way
(cf. Turner 1997). It has been suggested that such dark clouds in galaxies
like the SMC may in fact 
contain significant atomic components (e.g., Dickey 1996;
Kenney 1997). Even if such clouds are primarily molecular, their lower
densities imply long timescales for the formation of CO molecules, and
hence a reduction in the size of the CO-emitting regions within clouds
of a given radius (e.g., Maloney
1990). In this case, CO measurements will significantly underestimate
the molecular gas content of the clouds. Further uncertainties in the
empirical H$_{2}$ estimate come from possible metallicity dependence
on the CO-to-H$_{2}$ conversion factor (e.g., Maloney 1990).
A final possibility is that the gas-to-dust ratio
differs
between  diffuse and clumpy ISM regions regions. This could occur, for
example, if dust is accreted onto denser clouds, while dust in the
diffuse medium is destroyed by shocks on timescales much shorter than
new dust grains are created (e.g., Tielens 1998), although others have
argued that gas and dust should be generally well mixed in the ISM
(e.g., Cox \& Mezger 1989).

Regardless of the exact atomic versus molecular fraction in the clumpy
ISM of UGC~7321, its very existence has interesting implications for
furthering our understanding of the conditions in the ISM of LSB
galaxies.
Previously, it has been suggested that it is likely that 
interstellar pressures in LSB galaxies would be
too low to sustain a multi-phase structure in the diffuse,
low metallicity environments in LSB galaxies (Mihos, Spaans, \&
McGaugh 1999). Our data and models confirm that at least some LSB
galaxies can maintain sufficient pressures to support a
modest, multi-phase ISM.  In addition, the existence of dust in 
ISM regions of enhanced density
suggests that LSB galaxies have the envionments and the catalysts
needed for molecular hydrogen formation.

In general, a clumpy ISM should support higher H$_{2}$ fractions than
a nearly smooth ISM due
to the presence of more efficient shielding of molecules from
ionizing radiation fields (see also Spaans 1998). Thus even if the
dark clouds in UGC~7321 contain significant atomic components, their
inner regions almost certainly contain molecular cores (as confirmed
by the CO detection of Matthews \& Gao), hence eliminating the need to
invoke star formation in LSB galaxies directly from atomic gas
(cf. Schombert \etal\ 1990). This also 
raises the probability that other LSB galaxies could be
detected in CO with sufficiently deep integrations (see also
Matthews \& Gao 2000).

\section{Model Rotation Curves\protect\label{curves}}
\subsection{Background}
\subsubsection{The Importance of LSB Galaxy Rotation Curves}
Because of their relatively low visible matter contents
relative to their inferred dynamical masses,
LSB galaxies offer an important class of objects for
unravelling the nature of dark matter in galaxies, as well as for testing dark
matter alternatives, such as Modified Newtonian Dynamics (MOND; Milgrom
1983). An important step in this process is deriving accurate rotation
curves. 

For some applications,
systems that are nearly edge-on
can be particularly desirable for rotation curve studies
since inclination corrections to line-of-sight velocities
and position angle uncertainties in
these cases are small. \HII\ regions
can also in general be observed to larger galactocentric radii in
edge-on galaxies than in less highly
inclined systems, and edge-on galaxies are much easier to pick out
and observe at higher redshifts.

Unfortunately, the interpretation of observed rotation 
curves via mass-modelling techniques (e.g., Carignan 1985)
is strongly dependent on knowledge of the precise shape of the inner,
rising portion of the
rotation curve. This is often poorly determined
observationally, particularly in edge-on galaxies. If rotation
curves  are derived from
\HI\ data, they 
often suffer from beam smearing (see, e.g., 
Swaters 1999 and references therein; 
van den Bosch \etal\ 2000). High spatial resolution optical 
data can overcome this problem, but
instead suffer from the vagaries and uncertainties 
of internal absorption in a given galaxy, and the debate over
how accurately \HII\ regions trace the true disk
rotation (e.g., Davies 1991; Prada \etal\
1994; McKeith \etal\ 1995; 
Prada 1996). Uncertainties due to projection effects and
unknown gas distributions will also plague the interpretation of both
optical and \HI\ data in the edge-on case.
 
Our Monte Carlo simulations provide a means to
derive model optical rotation
curves for edge-on galaxies and to test the severity of some of 
the influences mentioned above. Bosma
\etal\ (1992) and Bosma (1995) have already shown via rotation curve models
and a 
comparison between \HI-, CO-, and
optically-derived rotation curves that even edge-on galaxies tend to
be largely optically thin.  Bosma
\etal\ (1992), Byun (1993), and Bosma (1995) 
also showed that extremely high optical depths are
needed to produce an apparent linear rotation curve from an intrinsically steeply
rising one, even in edge-on and nearly edge-on galaxies.  
Here we reconfirm their results via Monte Carlo techniques,
and also concentrate more
specifically on exploring situations relevant to the interpretation of
the rotation curve data for edge-on and nearly edge-on LSB
galaxies, including UGC~7321. We also assess the promise of using such
systems  for detailed dynamical studies and mass
modelling. 

\subsection{Artificial Rotation Curves: the Method}
For the 
present study, we model
galactic rotation curves by inputting an analytic  galactic
rotational law for our model galaxy, and then 
adding together the Doppler shifts 
that arise between the emitted photons (presumed to arise 
from H$\alpha$-emitting ``\HII\ regions'') 
and the observer. We found that also including the Doppler 
shifts arising from the relative bulk motions of the scattering dust 
particles had negligible effect on the final rotation curves
(see also Bosma \etal\ 1992). 
We assume that the dust opacity is
constant over the narrow wavelength range that an observer would use
to trace the galactic rotation.
Because we are keeping track of wavelength in these simulations, 
when a photon exits the galaxy 
%(either directly or by scattering) 
we project 
the photon into an image, and form ``channel maps'' by assigning an image to 
a particular wavelength range.  

The output of our radiation transfer code is 
a 3D data cube consisting of $N$ two-dimensional channel maps.  
To form a rotation curve from our channel maps we take a strip across the 
major axis of each channel map and calculate the flux-weighted velocity at 
each position along the strip
\begin{equation}
v(x)={ {\sum_{i=1}^{N} v_i F_i(x)} / {\sum_{i=1}^{N} F_i(x)} } \; ,
\end{equation}
where $x$ is the position along the ``slit'', $v_i$ is the velocity 
corresponding to the $i$th image channel, and $F_i(x)$ is the flux at 
position $x$ in the $i$th image channel. We note that this type of
intensity-weighted determination of $v(x)$ is technically correct
only for non-edge-on galaxies (e.g., Sancisi \& Allen 1979). 
However, it is still most
commonly the method employed by optical observers of edge-on galaxies
(e.g., Goad \& Roberts 1981; Karachentsev \& Xu 1991; Makarov,
Burenkov, \& Tyurina 1999; 
Dalcanton \& Bernstein 2000), hence we adopt it here for ease of
comparison of our models with data from the literature. The intrinsic
spectral resolution of our models is 4.5~\kms\ pixel$^{-1}$. Before 
plotting our model rotation curves, we convolved them with
Gaussians representing the expected turbulence of the ionized gas (assumed to
have $\sigma$=10~\kms) and a typical instrumental resolution (taken to
be 30~\kms). For simplicity, we begin by exploring smooth emissivity
distributions. The case of clumpy emissivity is explored in
Sect.~\ref{clumpcurve}. 

\subsection{Artificial Rotation Curves: the Models}
As discussed above,
the  shape of the rotation curve observed for an edge-on galaxy at
optical wavelengths may arise
from a combination of effects, including:  
(1) the intrinsic shape
of the underlying rotation curve; (2) projection effects; (3) dust
extinction; (4) anomalously high extinction in the center only;
(5) a complete  absence of ionized gas in the
central regions; (6) clumpy or irregularly distributed gas. Here explore the 
consequences of these effects for the observed rotation curves of edge-on
galaxies similar to UGC~7321.

\subsubsection{Effects of Internal Extinction, Projection, and
Intrinsic Rotation Curve Shape\protect\label{shapes}}
Let us first consider models that explore items (1)-(3) listed above, i.e.,
intrinsic rotation curve shape, projection effects, and wide-spread
dust extinction.

We take UGC~7321 as our template galaxy, adopting in all of our model
rotation curves
its inferred dust model distribution from Sect.~\ref{basicin} \& \ref{clumpin},
and a maximum rotational velocity of 100~\kms\ (MGvD99). For simplicity,
we assume the  scale height and scale length of the H$\alpha$ corresponds
to that of the  
stars.

We begin with the simple case of a smooth dust emissivity distribution 
and consider edge-on optical depths of
$\tau_{B,e}$=0,2.0, and 100.  We compute our models for three inclinations:
$i=85^{\circ}$,$88^{\circ}$, and 90$^{\circ}$. As already shown by Bosma
\etal\ (1992), the effects of items (1)-(3) above begin to become
negligible at inclinations $i\la 85^{\circ}$. 

For our input rotation laws, we choose four
different rotation curve shapes. Three of the forms are
based on the observed rotation curves
of edge-on galaxies in the literature having luminosities and peak rotational
velocities similar to UGC~7321. The adopted models are $(i)$ a very steeply 
rising curve, rising linearly to $V_{max}$ within 0.25~kpc and staying flat
thereafter;  $(ii)$ a moderately steep curve, rising linearly to $V_{max}$
within the inner 1~kpc and flattening thereafter 
(e.g., UGC~3697, Goad \& Roberts
1981); $(iii)$ an ``arctan'' rotation
curve (see Courteau 1998) of the form $v(R)=v_{0} +
\frac{2}{\pi}v_{c}{\rm arctan}(R)$ where $R=(r-r_{0})/r_{t}$, $v_{0}$
is the velocity center of rotation, $r_{0}$ is the spatial center of
the galaxy, $v_{c}$ is an asymptotic velocity, and $r_{t}$ is a
transition radius between the rising and flat parts of the rotation
curve (e.g., UGC~7321; Goad \& Roberts 1981). For the arctan model 
we adopt $v_{0}$=0, $v_{c}$=100~\kms,
$r_{0}$=0, and $r_{t}$=1.3~kpc. Finally
$(iv)$ is a rotation curve that rises
linearly to 2.2$h_{r}$  and has only an essentially negligible flat
portion within the stellar disk (see UGC~9242, Goad \& Roberts 1981;
FGC~1285, Dalcanton \& Bernstein 2000).

Our resulting models for these four different rotation curve shapes at three
inclinations and three different optical depths are illustrated in
Figure~\ref{fig:multirot}. In Figure~\ref{fig:goadplot} we also show
$\tau_{e,B}$=4.0 models for input rotation curve forms $(ii)$ and $(iii)$
with the H$\alpha$ emission line data
from Goad \& Roberts (1981) overplotted. 

Several interesting trends are immediately 
evident in Figure~\ref{fig:multirot}.  First, our model
curves reconfirm the result of Bosma \etal\ (1992) that an observed
rotation curve that is linear and rising to the outer regions of the
stellar disk cannot be produced by any physically plausible amount of
internal extinction. For $i=88^{\circ}$, 
we find that edge-on optical depths 
of $\tau_{B,e}\ga$100  are needed to
produce such an observed rotation curve for any type of intrinsic
rotation curve shape resembling models $(i)$-$(iii)$--i.e. for any
shape other than an intrinsically slowly rising one. Moreover, it becomes clear
from Figure~\ref{fig:multirot} that such a linear
shape cannot be solely the result of deriving the rotational velocities
via ``intensity-weighted'' procedures. We conclude that
edge-on
galaxies exhibiting linear, slowly rising rotation curves throughout
their stellar disks (e.g., UGC~9242, Goad \& Roberts 1981;
FGC~1285, Dalcanton \& Bernstein 2000) must necessarily be galaxies
with low central matter densities and are likely to generally
be optically thin systems. 
In the low optical depth regime appropriate for a typical
LSB  galaxy ($\bar\tau_{B,e}\sim$4; see Sect.~\ref{resultat}), the
effects of dust extinction on the observed rotation curve are  small enough
to be safely ignored even in the case of $i=90^{\circ}$.

At the opposite extreme, if a very steeply rising rotation curve that
flattens appreciably in the inner few kpc 
is directly observed in a highly inclined galaxy
(e.g., UGC~3697; Goad \& Roberts 1981), Figure~\ref{fig:multirot} 
shows that one can
infer that not only must the galaxy be relatively optically thin, but
it must also be inclined at an angle $i\la 85^{\circ}$ in order for
its steep rise to be preserved in projection. 

Figure~\ref{fig:goadplot} illustrates  that either of the two
intermediate models [$(ii)$ or $(iii)$] provides a roughly equally
good match to the
overall shape of the UGC~7321 rotation curve as measured by Goad \&
Roberts (1981), allowing for observational uncertainties and a
slight lopsidedness in the true rotation curve. Thus we see that 
in these intermediate
regimes, recovering the true, underlying rotation curve shape for a
particular galaxy is less
clear-cut. All of our models show that for the optical depths
appropriate for a galaxy like UGC~7321, the effects of extinction can
be safely ignored; instead, the greater ambiguity
arises from the fact that rotation curves of quite different 
shapes, steepnesses
and turnover radii can
look very similar in projection if the galaxy inclination angle
$i>85^{\circ}$. Hence, this could
create a much more serious problem for an observer attempting a dynamical
model of a galaxy like UGC~7321. In the case where the inclination is
not precisely known, the ambiguity becomes even more severe.

\subsubsection{Effects of Galaxy Centers Devoid of Gas}
It is sometimes suggested that the linear or slowly rising rotation
curves of some edge-on galaxies are due not to high optical depths
throughout the disk,
but rather a dearth of gaseous material in the inner regions of the
galaxy, or the possibility that the  \HII\ regions (or \HI\
gas) are confined primarily to a ring-like distribution. 
The former case is essentially equivalent to a galaxy which for
some reason may have a high optical depth near its center, 
but still may have an
optically thin outer disk (cf. Davies 1991). Using our models, we explore
the applicability of such models to UGC~7321, and to explaining
the linear, slowly rising rotation curves seen numerous late-type,
edge-on galaxies (e.g., Goad \& Roberts 1981;  
Karachentsev \& Xu 1991; Makarov \etal\ 1997;
Makarov, Burenkov, \& Tyurina 1999; Dalcanton \& Bernstein 2000).

We are immediately able to rule out a pure \HII\ region ``ring'' model
for UGC~7321, and indeed, for most late-type,
edge-on galaxies where H$\alpha$
imaging data are available. 
Such a ring morphology creates an H$\alpha$
intensity distribution with  peaks at the galactocentric
radii corresponding to the radius of the projected ring; such 
a distribution is inconsistent with H$\alpha$ imaging 
data 
available for UGC~7321 (MGvD99),
for several other superthin galaxies (R\"onnback \& Bergvall 1995; 
Hoopes, Walterbos, \& Rand 1999; Matthews \etal\ unpublished) and for
numerous less inclined LSB disk galaxies (e.g., Schombert \etal\ 1992;
O'Neil \etal\ 1998).

Although the ring model seems implausible, MGvD99 did report a
paucity of H$\alpha$ emission near the center of the UGC~7321 disk,
suggesting the possibility of a gas-poor or high optical
depth region in the inner 0.5~kpc or so of the galaxy. For illustration
purposes, here we test a
somewhat more extreme version of this model.
In Figure~\ref{fig:hole}, we show a model for $i=88^{\circ}$ and rotation curve
model $(ii)$, but this time the model galaxy is completely devoid of
H$\alpha$ emission within a central hole of radius 1.5~kpc. All other
parameters are as in the models in Figure~\ref{fig:multirot}.

It can be seen in
Figure~\ref{fig:hole} that the effect of this H$\alpha$-poor central disk
region  is to make the slope of the rotation
curve slightly shallower than the case where the emission extends all
the way to the center (cf. Figure~\ref{fig:multirot}). 
However, this is still not sufficient
to produce a linear rotation curve shape. The effect would of course
increase somewhat if the hole were made larger, but such a model 
seems highly contrived; we are aware of no examples of  
Sd spiral galaxies devoid of
H$\alpha$ emission over such a large fraction of the disk, 
nor should optical depth
effects completely prevent H$\alpha$ from being observable at small radii
in normal late-type spirals, particularly LSBs.
Hence this model should illustrate the upper limit of this effect
for a realistic Sd galaxy. We conclude that the uncertainty of how far
the emitting gas extends to the center of a galaxy will create
additional small
uncertainties in uncovering the the intrinsic inner rotation curve shape for
galaxies with $i\ge 85^{\circ}$. However this effect cannot in itself
produce an apparent
linearly rising rotation curve for physically realistic galaxy models.

\subsubsection{The Effects of Clumpy of Sparsely Distributed \HII\
Regions\protect\label{clumpcurve}}
Up until now, we have considered only smooth emissivity distributions
for our rotation curve models. However, the \HII\ regions from which a
rotation curve would be derived in a real galaxy are likely to have a
clumpy and somewhat irregular distribution. Their distribution may be
particularly sparse or irregular in the case of LSB galaxies
(cf. Schombert \etal\ 1992; O'Neil \etal\ 1998). To
explore how this may affect an observed rotation curve, we now compute a
model where the H$\alpha$ filling factor $ff_{H\alpha}$
is only 0.5\% and where $C_{H\alpha}$=100.

In Figure~\ref{fig:clumpHA} we show these ``clumpy'' models for the
$i=88^{\circ}$ case and input rotation curve form $(ii)$.
 As expected, the models now show a much more ragged
appearance, with many of the fluctuations similar in amplitude to those in the
real data. We see that an observed rotation curve with a
significant number of ``wiggles'' is not necessarily a signature of
strong patchy absorption, but could also occur in an extremely optically
thin galaxy only sparsely populated with \HII\ regions, as is not uncommon
in the case of LSB galaxies. Our models suggest that these fluctuations may
introduce some additional uncertainty in  recovering the precise slope
of the inner rotation curve of some LSB galaxies.

\section{Summary}
We have presented some of the first 3D Monte Carlo simulations
of the dusty ISM properties in the low optical depth regimes encountered
in low surface brightness (LSB) spiral galaxies. We have demonstrated
the power of using such techniques in combination with
multiwavelength observational data to constrain the 
amount and distribution of dust in such systems. Our realistic models
fully take into account scattering and the  effects of a clumpy,
multi-phase ISM.

Because late-type
LSB galaxies are relatively optically thin, edge-on examples of these
systems allow a unique opportunity to explore the
structure and distribution of their ISM and place
important empirical constraints on dust models.
We have shown that while dust contents in LSB galaxies are relatively
low, these galaxies can contain modest amounts of dust and molecular
material. Moreover, in at least some cases,
a significant fraction of this dusty material
is contained within a clumpy medium, confirming that
even LSB galaxies can maintain sufficient pressures to
support a modest multi-phase ISM structure.

From detailed models of the nearly edge-on LSB
galaxy UGC~7321 ($i=88^{\circ}$), our Monte Carlo
models indicate an edge-on optical
depth $\bar\tau_{e,B}\sim$4.0, and hence a total mass of dust 
is this galaxy of $\sim 8.8\times10^{5}$~\msun. This is
$\sim20\times$ the warm dust content inferred from {\it IRAS}
100$\mu$m measurements.  Total
$B$-band extinction toward the center of this galaxy is estimated to
be $\sim0.43\pm0.2$ mag.

We infer that $\sim$50\% of the dusty material in UGC~7321 is
contained in a clumpy medium. Based on the detection of CO emission
from UGC~7321 by Matthews \& Gao (2000), it appears that some
fraction of the clumped material is almost certainly molecular
gas. However, the dark clouds in LSB galaxies may also contain
significant fractions of atomic hydrogen gas. Together our models and
direct \HI\ and CO measurements  suggest that UGC~7321 has
a gas-to-dust ratio of at least 850 in the inner regions of its
stellar disk, while the global gas-to-dust ratio for the entire galaxy
(including \HI\ gas extending beyond the stellar disk) 
is estimated to be $\sim$1600.

In spite of the modest dust contents of LSB galaxies, we have shown that
for the range of optical depths expected for such systems, dust
extinction will have appreciable effects on the observed total magnitudes
and colors only when these galaxies are observed near edge-on. However, even 
then, the amount of reddening due to dust will be insufficient to explain
the large  radial color gradients now observed in a number of 
LSB galaxies (de Blok \etal\ 1996; MGvD99; Matthews \etal\ 2000;
Bell \etal\ 2000). Thus
some LSB galaxies must have large intrinsic stellar population and/or
metallicity  gradients in their disks.

We have shown that for realistic optical depths, dust has no appreciable
effect on the  rotation curves of edge-on spiral galaxies observed
in the H$\alpha$ emission line.  Possible holes in the central  H$\alpha$
distributions of such galaxies also cannot fully account for
the observed slowly rising rotation curves. Thus the linearly, slowly 
rising rotation
curves seen frequently in many  late-type, edge-on or
nearly edge-on  spirals
cannot be fully accounted for by dust and must be due to low central
matter densities in these systems.  We have demonstrated that
in general, projection effects will create a far greater uncertainty 
than optical depth effects in interpreting
the precise intrinsic shapes of the rotation curves of edge-on galaxies.

\acknowledgements
We thank Mike Wolff for valuable discussions during the course of this
work, and 
KW also thanks Elizabeth Barton and Margaret Geller for many discussions on 
galaxy rotation curves during the development of the Monte Carlo radiation 
transfer code that simulates the scattering of emission lines in 
moving media.  
LDM is grateful for the support  provided by a
Jansky Postdoctoral Fellowship from the National Radio Astronomy
Observatory.
KW acknowledges support from NASA's Long Term Space Astrophysics Research 
Program (NAG5-6039). The WFPC2 imaging data used for this work
were obtained as
part of the Wide Field and Planetary Camera~2 Investigation Definition
Team science programs.  This publication also made use of data products
from the Two Micron All Sky Survey, which is a joint project of the
University of
Massachusetts and the Infrared Processing and Analysis
Center/California
Institute of Technology, funded by the National
Aeronautics and Space Administration and the National Science Foundation.

\clearpage

%Fig. 1
\figcaption{$F702W$+$F814W$ ($R+I$) composite image of a
portion of the disk of UGC~7321, as imaged by the Wide Field and
Planetary Camera~2 (WFPC2) aboard the {\it Hubble Space Telescope}.
This is an 55$'' \times 19''$ section of the disk, centered near
$r=0$. Resolution is $\sim$\as{0}{1}. 
White areas reveal the presence of clumpy,
optically thick material in the disk of this LSB 
galaxy.\protect\label{fig:WFPC2}}

%Fig. 2
\figcaption{Observed radial $B-R$ color gradient  along the major
axis of UGC~7321. For ease of comparison with the model
computations presented in this paper, the color has been shifted 
by a constant $c_{1}$ such that $(B-R)+ c_{1}\approx$0 at the last
reliably measured data points, near
$r=$6~kpc. The data were extracted by averaging over
 a 12-pixel (\as{2}{3}) wide strip, and smoothed by a factor of 15 for
display purposes.
For further details see MGvD99.\protect\label{fig:BRreal}}

%Fig. 3
\figcaption{Observed $R-H$ color gradient along the major axis of UGC~7321.
The data were extracted by averaging 
along an 3$''$ wide strip and then smoothed by a 
factor of 3 for display purposes. 
For ease of comparison with model calculations presented
in this paper, the color has been shifted by a constant factor $c_{2}$
such that
$(R-H)+c_{2}\approx$0 near the last reliable data points at 
$r=\pm 3.5$~kpc.\protect\label{fig:RHreal}}

%Fig. 4
\figcaption{$R$-band scattered light images generated from our Monte
Carlo simulations of UGC~7321 at $i=88^{\circ}$. Each
panel is 12.3~kpc is diameter, corresponding to the 
disk region over which our models were
computed. The lefthand column shows models
with smooth dust and emissivity distributions, and the righthand
column shows models with a 2-phase clumpy ISM and 
stellar components. Further details on the input parameters for the models
are summarized in Sect.~\ref{smooth} \& \ref{clumpin} and 
Table~2. The four panels in each column represent
four different edge-on $B$-band optical depths: $\tau_{e,B}$=8.0
(top); 4.0 (second row); 2.0 (third row); and 0.4 
(bottom).\protect\label{fig:Rscat}}

%Fig. 5
\figcaption{Comparison of our best $R$- and $H$-band
Monte Carlo model images of UGC~7321 with real data at these wavebands.
Top: an $R$-band image of UGC~7321 obtained with the
WIYN telescope by MGvD99 and degraded in resolution to match the
resolution of our models ($\sim$60~pc pixel$^{-1}$. Only the inner 12.3~kpc
of the galaxy is shown. Row 2: an $R$-band model image with $\tau_{e,B}$=4.0
and clumpy ISM and emissivity distributions. Third row:
an $H$-band image of the inner 12.3~kpc of UGC~7321 obtained by MGvD99. Bottom:
our model $H$-band image with $\tau_{e,B}$=4.0
and clumpy ISM and emissivity distributions.\protect\label{fig:zoom}}

%Fig. 6
\figcaption{Differential $B-R$ and $R-H$ color profiles
along the galaxy major axis for our family of
$i=88^{\circ}$ smooth ISM models (upper panels; see Sect.~\ref{smooth}) and 
our clumpy ISM models (lower panels; see Sect.~\ref{clumpgrad}).  
Shown are the color difference between
models of 4 different edge-on 
optical depths $\tau_{e,B}$ relative to a zero dust model, as a
function of galactocentric distance: $\tau_{e,B}=8.0$ (solid line);
$\tau_{e,B}=4.0$ (dotted line); 
$\tau_{e,B}=2.0$ (dashed line); $\tau_{e,B}=0.4$
(dot-dash line).\protect\label{fig:smoothcolor}}

%Fig. 7
\figcaption{Illustration of the origin of the $\Delta(B-R)$ 
``saturation'' effect
for a slab model of uniform density and emissivity. 
Plotted are
the color differences in $B-R$ and $R-H$  
between a constant density absorbing
medium with no scattering and uniform emissivity, and a uniform dust-free
emitting medium, as a function of $B$-band edge-on optical depth $\tau_{B}$
through the slab (see Equation~4).
It can be seen that $\Delta(B-R)$ reaches a constant value at a
considerably lower optical depth than $\Delta(R-H)$.\protect\label{fig:sat}}

%Fig. 8
\figcaption{Differential $H-K$ color profiles
along the galaxy major axis for our family of
$i=88^{\circ}$ clumpy models.  Shown are the color difference between
models of 4 different optical depths relative to a zero dust model, as a
function of galactocentric distance. The 4 different line styles correspond to
the same edge-on optical depths as in Figure~\ref{fig:smoothcolor}. 
\protect\label{fig:HKmodcolor}}

%Fig 9
\figcaption{$H-K_{s}$ color profile along the major axis of UGC~7321, derived
from data from the Two Micron All-Sky Survey (2MASS). The data were 
extracted by averaging over a 3-pixel wide ($\sim 3''$) strip. The
signal-to-noise of the data deteriorate significantly outside the central kpc
of the galaxy (see Jarrett \etal\ 2000), 
but they are consistent with no significant $H-K_{s}$ 
color gradient.
For ease of comparison with model calculations presented
in this paper, the observed 
color has been shifted by a constant factor $c_{3}$
such that $(H-K_{s})+c_{3}\approx$0.\protect\label{fig:HKreal}}

%Fig. 10
\figcaption{Dust-induced vertical
$B-R$ color gradients, extracted from our clumpy
models parallel to the minor
axis at three different disk positions. 
Shown are the color difference between
models of 4 different edge-on 
optical depths $\tau_{e,B}$ relative to a zero dust model, as a
function of galactocentric distance: $\tau_{e,B}=8.0$ (solid line);
$\tau_{e,B}=4.0$ (dotted line); 
$\tau_{e,B}=2.0$ (dashed line); $\tau_{e,B}=0.4$
(dot-dash line).
The top panel shows the minor axis color
profile; the middle panel shows the color at $r$=1.5~kpc, and the
bottom panel along $r=3.0$~kpc.\protect\label{fig:fakevert}}

%Fig. 11
\figcaption{Observed  $B-R$ vertical 
color profiles as a function of $z$ (in kpc)
for UGC~7321. The data were  extracted over 15-pixel
($\sim$\as{3}{0}) wide strips
near the minor axis (top) and at $r$=3.0~kpc (bottom). For display
purposes the data were smoothed by a factor of 3. For ease of
comparison with the models in Figure~\ref{fig:fakevert}, 
the data have been shifted by a
constant factor $c_{4}$ such that $(B-R)+c_{4}\approx$0 
near the last reliably measured data points, at 
$r=\pm$0.4~kpc.\protect\label{fig:realvert}}

%Fig. 12
\figcaption{$B$-, 
$R$-, $H$-band normalized major axis intensity profiles for our clumpy models
viewed at $i=88^{\circ}$, with
$\bar\tau_{e,B}$=0 (thick solid line);
0.4 (dot-dash line); 2.0 (dashed line); 4.0 (dotted line); and 8.0 
(solid line).\protect\label{fig:radprofs}}

%Fig. 13
\figcaption{Dust-induced $B-R$ and $R-H$ color gradients as a function
of galactocentric radius (in kpc) for smooth models 
models viewed at inclinations of $i=0^{\circ}$, 
45$^{\circ}$, \& 90$^{\circ}$. For ease of comparison,
the $i=88^{\circ}$ smooth models from Figure~\ref{fig:smoothcolor} are also
reproduced here. Shown are the color difference between
models of 4 different edge-on 
optical depths $\tau_{e,B}$ relative to a zero dust model, as a
function of galactocentric distance: $\tau_{e,B}=8.0$ (solid line);
$\tau_{e,B}=4.0$ (dotted line); 
$\tau_{e,B}=2.0$ (dashed line); $\tau_{e,B}=0.4$
(dot-dash line).\protect\label{fig:lowincsm}}

%Fig. 14
\figcaption{Dust-induced $B-R$ and $R-H$ color gradients as a function
of galactocentric radius (in kpc) for clumpy models 
models viewed at inclinations of $i=0^{\circ}$, 
45$^{\circ}$, \& 90$^{\circ}$. Line styles are as in
Figure~\ref{fig:lowincsm}. For ease of comparison,
the $i=88^{\circ}$ clumpy models from Figure~\ref{fig:smoothcolor} are also
reproduced here.\protect\label{fig:lowinccl}}

%Fig. 15
\figcaption{Rotation curves generated from
our Monte Carlo models for 3 values of
$\tau_{e,B}$ (0, 2, \& 100)
and 3 viewing angles ($i=85^{\circ}$,88$^{\circ}$, \& 90$^{\circ}$) 
for the four analytic rotation curve
models described in Sect.~\ref{shapes}. Each column shows a different input
model. The thick solid line in each panel shows the analytic
form of the input rotation curve without any internal
extinction. The thin solid line shows the ``observed'' intensity-weighted
rotation curve for the $\tau_{e,B}$=0 (zero dust) 
model. The dotted line shows the
predicted observed curve for a $\tau_{e,B}$=2 model, and the dashed line the
predicted observed curve for a
$\tau_{e,B}$=100 model. In general, the $\tau_{e,B}$=0 and
$\tau_{e,B}$=2 models are nearly
indistinguishable.\protect\label{fig:multirot}}

%Fig. 16
\figcaption{Rotation curves generated from our Monte Carlo models for
$\tau_{e,B}$=4.0 and a viewing angle of $i=88^{\circ}$. The thick
solid line shows the form of the input rotation curve
[model $(ii)$ (top) and model $(iii)$ (bottom; see Text)], and
the thin solid line shows the predicted observed rotation curve.
Overplotted
are datapoints (triangles) measured for UGC~7321 via H$\alpha$ emission line
spectroscopy by Goad \& Roberts (1981).\protect\label{fig:goadplot}}

%Fig. 17
\figcaption{Model rotation curves generated for a galaxy seen at
$i=88^{\circ}$ and devoid of H$\alpha$ emission over a central hole of
radius 1.5~kpc. The thick solid line shows the 
form of the input rotation curve without any internal
extinction. The thin solid line shows the ``observed'' intensity-weighted
rotation curve for the $\tau_{e,B}$=0 (zero dust) 
model. The dotted line shows the
predicted observed curve for a $\tau_{e,B}$=2 model, and the dashed line the
predicted observed curve for a
$\tau_{e,B}$=100 model.
Overplotted
are data points (triangles) measured for UGC~7321 by Goad \& Roberts
(1981) via H$\alpha$ emission line
spectroscopy.\protect\label{fig:hole} }

%Fig. 18
\figcaption{Model rotation curves generated for a galaxy seen at
$i=88^{\circ}$ and having a clumpy distribution of \HII\ regions, with
a filling factor $ff_{H\alpha}$=0.005.  The thick solid line shows 
the analytic
form of the input rotation curve without any internal
extinction. The thin solid line shows the ``observed'' intensity-weighted
rotation curve for the $\tau_{e,B}$=0 (zero dust) 
model. The dotted line shows the
predicted observed curve for a $\tau_{e,B}$=2 model, and the dashed line the
predicted observed curve for a
$\tau_{e,B}$=100 model. Overplotted
are data points (triangles) measured for UGC~7321 via H$\alpha$ emission line
spectroscopy by Goad \& Roberts (1981).\protect\label{fig:clumpHA} }


\begin{references}
Bell, E. F., Barnaby, D., Bower, R. G., de Jong, R. S., Harper,
D. A. Jr., Hereld, M., Loewenstein, R. F., \& Rauscher, B. J. 2000,
MNRAS, 312, 470

Bergvall, N. \& R\"onnback, J. 1995, MNRAS, 273, 603

Bergvall, N., R\"onnback, J., Masegosa, J., \"Ostlin, G. 1999, A\&A,
341, 697

Bianchi, S., Ferrara, A., Davies, J. I., \& Alton, R. B. 2000, MNRAS,
311, 601

Bianchi, S., Ferrara, A., \& Giovanardi, C. 1996, ApJ, 465, 127

Block, D. L., Witt, A. N., Gosb{\o}l, P., Stockton, A., \& Moneti,
A. 1994, A\&A, 288, 383

Boiss\'e, P. 1990, A\&A, 228, 483

Bosma, A. 1995, in The Opacity of Spiral Disks, edited by J. I. Davies
and D. Burstein, (Dordrecht: Kluwer), p. 317

Bosma, A., Byun, Y., Freeman, K. C., \& Athanassoula, E. 1992, ApJ,
400, L21

Bothun, G., Impey, C., \& McGaugh, S. 1997, PASP, 109, 745

Bouchet, P., Lequeux, J., Maurice, E., Pr\'evot, L., \&
Pr\'evot-Burnichon, M. L. 1985, A\&A, 149, 330

Byun, Y.-I. 1993, PASP, 105, 993

%Byun, Y. I., Freeman, K. C., \& Kylafis, N. D. 1994, ApJ, 432, 114

Carignan, C. 1985, ApJ, 299, 59

Code, A. D. \& Whitney, B. A. 1995, ApJ, 441, 400

Cole, A. A., Wood, K., \& Nordsieck, K. H. 1999, AJ, 118, 2292

Courteau, S. 1998, AJ, 114, 2402

Cox, P. \& Mezger, P. G. 1989, A\&ARv, 1, 49

Dalcanton, J. J. \& Bernstein, R. A. 2000, in Dynamics of Galaxies:
from the Early Universe to the Present, edited by F. COmbes,
G. A. Mamon, \& V. Charmandaris, (San Francisco: ASP), 161

Dalcanton, J. J. \& Schectman, S. A. 1996, ApJ, 465, L9

Davies, J. I. 1991, in Dynamics of Disc Galaxies, edited by
B. Sundelius, (G\"oteborg: G\"oteborgs University and Chalmers
University of Technology), 65

de Blok, W. J. G. \& McGaugh, S. S. 1997, MNRAS, 290, 533

de Blok, W. J. G. \& van der Hulst, J. M. 1998, A\&A, 336, 49

de Blok, W. J. G., van der Hulst, J. M., \& Bothun, G. D. 1995, MNRAS,
274, 235

de Grijs, R. 1998, MNRAS, 299, 595

de Jong, R. S. 1996, A\&A, 313, 377

Devereux, N. \& Young, J. S. 1990, ApJ, 359, 42

Dickey, J. M. 1996, in The Minnesota Lectures on Extragalactic
Hydrogen, edited by E. D. Skillman, (San Francisco: ASP), 187

Disney, M., Davies, J., \& Phillipps, S. 1989, MNRAS, 239, 939

Dwek, E. 1998, ApJ, 501, 643

Elmegreen, B. G. 1993, in Protostars and Planets III, edited by
E. H. Levy and J. I. Lunine, (Tucson: University of Arizona Press), 97

Evans, R. 1994, MNRAS, 266, 511

Ferrara, A., Bianchi, S., Cimatti, A., \& Giovanardi, C. 1999, ApJS,
123, 437

Ferrara, A., Bianchi, S., Dettmar, R. -J., \& Giovanardi, C. 1996,
ApJ, 467, 69

Firmani, C., Hern\'andez, X., \& Gallagher, J. S. 1996, A\&A, 308, 403

Freeman, K. C. 1970, ApJ, 160, 811

Gallagher, J. S. \etal\ 2000, in prep.

Gerritsen, J. P. E. \& de Blok, W. J. G. 1999, A\&A, 342, 655

Goad, J. W. \& Roberts, M. S. 1981, ApJ, 250, 79

Greenberg, J. M. \& Li, A. 1995, in The Opacity of Spiral Disks,
edited by J. I. Davies and D. Burstein, (Dordrecht: Kluwer), 19

Han, M. 1992, ApJ, 391, 617

Henyey, L. G. \& Greenstein, J. L. 1941, ApJ, 93, 70

Hoeppe, G., Brinks, E., Klein, U., Giovanardi, C., Altschuler, D. R.,
Price, R. M., \& Deeg, H.-J. 1994, AJ, 108, 446

Hoopes, C. G., Walterbos, R. A. M., \& Rand, R. J. 1999, ApJ, 522, 669

Howk, J. C. \& Savage, B. D. 1999, AJ, 117,  2077

Hunter, D. A., Gallagher, J. S., Rice, W. L., \& Gillett, F. C. 1989,
ApJ, 336, 152

Impey, C. \& Bothun, G. 1997, ARA\&A, 35, 267

Jarrett, T. H., Chester, T., Cutri, R., Schneider, S., Skrutskie, M.,
\& Huchra, J. P. 2000, AJ, 119, 2498

Just, A., Fuchs, B., \& Wielen, R. 1996, A\&A, 309, 715

Karachentsev, I. D., Karachentseva, V. E., \& Parnovsky, S. L. 1993,
Astron. Nachr., 314, 97

Karachentsev, I. D. \& Xu, Z. 1991, Sov. Astron. Lett., 17, 135

Kenney, J. 1997, in The Interstellar Medium of Galaxies, edited by
J. M. van der Hulst, (Dordrecht: Kluwer), 33

Kent, S. M. 1986, AJ, 91, 1301

Kim, S. -H., Martin, P. G., \& Hendry, P. D. 1994, ApJ, 422, 164

Knezek, P. M. 1993, Ph.D. Thesis, University of Massachusetts

Koorneef, J. 1982, A\&A, 107, 247

Kravtsov, A. V., Klypin, A. A., Bullock, J. S., \& Primrack,
J. R. 1998, ApJ, 502, 48

Kuchinski, L. E. \& Terndrup, D. M. 1996, AJ, 111, 1073

Kuchinski, L. E., Terndrup, D. M., Gordon, K. D., \& Witt, A. N. 1998,
AJ, 115, 1438

Kylafis, N. D. \& Bahcall, J. N. 1987, ApJ, 317, 637

Lisenfeld, U. \& Ferrara, A. 1998, ApJ, 496, 145

Maloney, P. 1990, in The Interstellar Medium of Galaxies, edited by
H. A. Thronson, Jr. \& J. M. Shull, (Dordrecht: Kluwer), 493

Makarov, D. I., Burenkov, A. N., \& Tyurina, N. V. 1999,
Astron. Lett., 25, 706

Makarov, D. I., Karachentsev, I. D., Burenkov, A. N., Tyurina, N. V.,
\& Korotkova, G. G. 1997, Astron. Lett., 23, 638

Mathis, J. S., Rumpl, W., \& Nordsieck, K. H. 1977, ApJ, 217, 425

Matthews, L. D. 1998, Ph.D. Thesis, State University of New York at
Stony Brook

Matthews, L. D. 2000, AJ, in press

Matthews, L. D. \etal\ 1999, AJ, 118, 208

Matthews, L. D. \& Gallagher, J. S. 1997, AJ, 114, 1899

Matthews, L. D., Gallagher, J. S., \& van Driel, W. 1999,  AJ, 118,
2751 (MGvD99)

Matthews, L. D., Gallagher, J. S., \& van Driel, W. 2000, in 
Galaxy Dynamics: from the Early Universe to the Present, edited by
F. Combes, G. A. Mamon, \& V. Charmandaris, ASP Conference Series,
Vol. 197, (San Francisco: ASP), 195

Matthews, L. D. \& Gao. Y. 2000, in prep.

Matthews, L. D. \& van Driel, W. 2000, A\&AS, 143, 421

Matthews, L. D., van Driel, W., \& Gallagher, J. S. 1998, AJ, 116, 2196

McGaugh, S. S. 1994, ApJ, 426, 135

McGaugh, S. S. \& de Blok, W. J. G. 1998, ApJ, 499, 41

McGaugh, S. S., Schombert, J. M., Bothun, G. D., \& de Blok,
W. J. G. 2000, ApJ, 533, 99

McKeith, C. D., Greve, A., Downes, D., \& Prada, F. 1995, in The
Opacity of Spiral Disks, edited by J. I. Davies and D. Burstein,
(Dordrecht: Kluwer), 325

Mihos, J. C., Spaans, M., \& McGaugh, S. S. 1999, ApJ, 515, 89

Milgrom, M. 1983, ApJ, 270, 371

Misiriotis, A., Kylafis, N. D., Papamastorakis, J., \& Xilouris,
E. M. 2000, A\&A, 353, 117

O'Neil, K., Bothun, G. D., Impey, C. D., \& McGaugh, S. S. 1998, AJ,
116, 657

Pickering, T. E. \& van der Hulst, J. M. 1999, AAS, 195, 104.10

Pierni, D. 1999, A\&A, 352, 49

Prada, F. 1996, PASP, 108, 549

Prada, F., Beckman, J. E., McKeith, C. D., Castles, J., \& Greve,
A. 1994, ApJ, 423, L35

Rhee, M.-H. 1996, Ph.D. Thesis, University of Groningen

R\"onnback, J. \& Bergvall, N. 1995, A\&A, 302, 353

Sage, L. J. 1993, A\&A, 272, 123

Sancisi, R. \& Allen, R. J. 1979, A\&A, 74, 73

Sandage, A. \& Bedke, J. 1994, The Carnegie Atlas of Galaxies,
(Washington: Carnegie Institute of Washington)

Schombert, J. M., Bothun, G. D., Impey, C. D. \& Mundy, L. G. 1990,
AJ, 100, 1523

Schombert, J. M., Bothun, G. D., Schneider, S. E., \& McGaugh,
S. S. 1992, AJ, 103, 1107

Spaans, M. 1999, in The Low Surface Brightness Universe, edited by
J. I. Davies, C. Impey, \& S. Phillips, (San Francisco: ASP), 237

Sprayberry, D. Bernstein, G. M., Impey, C. D., \& Bothun, G. D. 1995,
ApJ, 438, 72

Stanimirovic, S., Staveley-Smith, L., van der Hulst, J. M., Bontekoe,
T. R., Kester, D. J. M., \& Jones, P. A. 2000, MNRAS, 315, 791

Stil, J. 1999, Ph.D. Thesis, University of Groningen

Swaters, R. A. 1999, Ph.D. Thesis, University of Groningen

Swaters, R. A., Madore, B. F., \& Trewhella, M. 2000, ApJ, 531, L107

Thronson, H. A. \& Telesco, C. M. 1986, ApJ, 311, 98

Tielens, A. G. G. M. 1998, ApJ, 499, 267

Trewehella, M., Madore, B., \& Kuchinski, L. 1999, in Observational
Cosmology: The Development of Galaxy Systems, ASP Conference Series,
Vol. 176 (San Francisco: ASP), 454

Tully, R. B., Pierce, M. J., Huang, J.-S., Saunders, W., Verheijen,
M. A. W., \& Witchalls, P. L. 1998, AJ, 115, 2264

Turner, B. E. 1997, in Astrophysical Implications of the Laboratory
Study of Presolar Materials, edited by T. J. Bernatowicz and
E. K. Zinner, (Woodbury: American Institute of Physics), 477

Turner, B. E. \& Ziurys, L. M. 1988,  in 
Galactic and Extragalactic Radio Astronomy,
edited by G. L. Vershuur and K. I. Kellermann, (New York:
Springer-Verlag), 200

van den Bosch, F. C. 2000, ApJ, 530, 177

van den Bosch, F. C., Robertson, B. E., Dalcanton, J. J., \& de Blok,
W. J. G. 2000, AJ, 119, 1579

Wainscoat, R. J., Freeman, K. C., \& Hyland, A. R. 1989, ApJ, 337, 163

Whittet, D. C. B. 1992, Dust in the Galactic Environment, (Bristol:
IOP Publishing)

Witt, A. N. 1977, ApJS, 35, 1

Witt, A. N. \& Gordon, K. D. 1996, ApJ, 463, 681

Witt, A. N. \& Gordon, K. D. 2000, ApJ, 528, 799

Witt, A. N., Oliveri, M. V., \& Schild, R. E. 1990, AJ, 99, 888

Wood, K. 1997, ApJ, 477, L25

Wood, K. \& Jones, T. J. 1997, AJ, 114, 1405

Wood, K. \& Reynolds, R. J. 1999, ApJ, 525,  799

Xilouris, E. M., Alton, P. B., Davies, J. I., Kylafis, N. D.,
Papamastorakis, J., \& Trewhella, M. 1998, A\&A, 331, 894

Xilouris, E. M., Byun, Y. I., Kylafis, N. D., Paleologou, E. V., \&
Papamastorakis, J. 1999, A\&A, 344, 868

Xilouris, E. M., Kylafis, N. D., Papamastorakis, J., Paleologou,
E. V., \& Haerendel, G. 1997, A\&A, 325, 135

Yusef-Zadeh, F., Morris, M., \& White, R. L. 1984, ApJ, 278, 186

Zwaan, M. A., van der Hulst, J. M., de Blok, W. J. G., \& McGaugh,
S. S. 1995, MNRAS, 273, L35

\end{references}
\end{document}